\documentclass[11pt,a4paper]{article} 

\usepackage[bindingoffset=0.3cm,textheight=22.5cm,hdivide={2.7cm,*,2.7cm}, vdivide={*,22cm,*}]{geometry}

\usepackage{setspace,graphics}
\usepackage[dvips]{epsfig}
\usepackage{amsmath,amscd,amssymb,epsfig,euscript,array,cite}
\usepackage{graphicx}
\usepackage{hyperref}
\usepackage[utf8]{inputenc}
\usepackage{fancyhdr}

\numberwithin{equation}{section} 

\newcounter{multieqs}

\newcommand{\ww}{\wedge}
\newcommand{\dd}{\ensuremath\mathrm{d}} 

\newcommand{\be}{\begin{equation}}
\newcommand{\ee}{\end{equation}}
\newcommand{\eq}[1]{(\ref{#1})}

\newcommand{\com}[2]{[ #1 , #2 ]}

\newcommand{\bm}[1]{\mbox{\boldmath $#1$}}

\def\bd{\begin{document}}
\def\ed{\end{document}}
\def\nn{\nonumber}
\def\bea{\begin{eqnarray}}
\def\eea{\end{eqnarray}}
\def\obar{\overline}
\let\bm=\bibitem

\newcommand{\EQ}[1]{\begin{equation} #1 \end{equation}}
\newcommand{\AL}[1]{\begin{subequations}\begin{align} #1 
\end{align}\end{subequations}}
\newcommand{\SP}[1]{\begin{equation}\begin{split} #1 \end{split}\end{equation}}
\newcommand{\ALAT}[2]{\begin{subequations}\begin{alignat}{#1} #2 
\end{alignat}\end{subequations}}
\def\beqa{\begin{eqnarray}} 
\def\eeqa{\end{eqnarray}} 
\def\beq{\begin{equation}} 
\def\eeq{\end{equation}} 

\def\N{{\cal N}}
\def\cJ{{\cal J}}
\def\sst{\scriptscriptstyle}
\def\thetabar{\bar\theta}
\def\Tr{{\rm Tr}}

\def\a{\alpha}          \def\da{{\dot\alpha}}
\def\b{\beta}           \def\db{{\dot\beta}}
\def\c{\gamma}  \def\C{\Gamma}  \def\cdt{\dot\gamma}
\def\d{\delta}  \def\D{\Delta}  \def\ddt{\dot\delta}
\def\e{\epsilon}                \def\vare{\varepsilon}
\def\f{\phi}    \def\F{\Phi}    \def\vvf{\f}
\def\g{\gamma} 
\def\h{\eta}
\def\k{\kappa}
\def\l{\lambda} \def\L{\Lambda} \def\la{\lambda}
\def\m{\mu}     \def\n{\nu}
\def\o{\omega}
\def\p{\pi}     \def\P{\Pi}
\def\r{\rho}
\def\s{\sigma}  \def\S{\Sigma}
\def\t{\tau}
\def\th{\theta} \def\Th{\Theta} \def\vth{\vartheta}
\def\X{\Xeta}
\def\z{\zeta}

\def\cA{{\cal A}} \def\cB{{\cal B}} \def\cC{{\cal C}}
\def\cD{{\cal D}} \def\cE{{\cal E}} \def\cF{{\cal F}}
\def\cG{{\cal G}} \def\cH{{\cal H}} \def\cI{{\cal I}}
\def\cJ{{\cal J}} \def\cK{{\cal K}} \def\cL{{\cal L}}
\def\cM{{\cal M}} \def\cN{{\cal N}} \def\cO{{\cal O}}
\def\cP{{\cal P}} \def\cQ{{\cal Q}} \def\cR{{\cal R}}
\def\cS{{\cal S}} \def\cT{{\cal T}} \def\cU{{\cal U}}
\def\cV{{\cal V}} \def\cW{{\cal W}} \def\cX{{\cal X}}
\def\cY{{\cal Y}} \def\cZ{{\cal Z}}

\def\mg{\mathfrak{g}}
\def\mh{\mathfrak{h}}
\def\mr{\mathfrak{r}}
\def\mt{\mathfrak{t}}

\def\und{\underline}
\def\ua{\underline{\alpha}}
\def\ub{\underline{\phantom{\alpha}}\!\!\!\beta}
\def\uc{\underline{\phantom{\alpha}}\!\!\!\gamma}
\def\um{\underline{\mu}}
\def\ud{\underline\delta}
\def\ue{\underline\epsilon}
\def\una{\underline a}\def\unA{\underline A}
\def\unb{\underline b}\def\unB{\underline B}
\def\unc{\underline c}\def\unC{\underline C}
\def\une{\underline e}\def\unE{\underline E}
\def\unf{\underline{\phantom{e}}\!\!\!\! f}\def\unF{\underline F}
\def\unm{\underline m}\def\unM{\underline M}
\def\unn{\underline n}\def\unN{\underline N}
\def\unp{\underline{\phantom{a}}\!\!\! p}\def\unP{\underline P}
\def\unq{\underline{\phantom{a}}\!\!\! q}
\def\unQ{\underline{\phantom{A}}\!\!\!\! Q}
\def\unH{\underline{H}}

\def\As {{A \hspace{-6.4pt} \slash}\;}
\def\bs {{b \hspace{-6.4pt} \slash}\;}
\def\Ds {{D \hspace{-6.4pt} \slash}\;}
\def\ds {{\del \hspace{-6.4pt} \slash}\;}
\def\ss {{\s \hspace{-6.4pt} \slash}\;}
\def\ks {{ k \hspace{-6.4pt} \slash}\;}
\def\ps {{p \hspace{-6.4pt} \slash}\;}
\def\pas {{{p_1} \hspace{-6.4pt} \slash}\;}
\def\pbs {{{p_2} \hspace{-6.4pt} \slash}\;}

\def\Fh{\hat{F}}
\def\Vh{\hat{V}}
\def\Xh{\hat{X}}
\def\ah{\hat{a}}
\def\xh{\hat{x}}
\def\yh{\hat{y}}
\def\ph{\hat{p}}
\def\xih{\hat{\xi}}

\def\psit{\tilde{\psi}}
\def\Psit{\tilde{\Psi}}
\def\tht{\tilde{\th}}
 
\def\At{\tilde{A}}
\def\Qt{\tilde{Q}}
\def\Rt{\tilde{R}}

\def\st{\tilde{s}}
\def\ft{\tilde{f}}
\def\pt{\tilde{p}}
\def\qt{\tilde{q}}
\def\vt{\tilde{v}}

\def\delb{\bar{\partial}}
\def\bz{\bar{z}}
\def\bD{\bar{D}}
\def\bB{\bar{B}}

\def\bk{{\bf k}}
\def\bl{{\bf l}}
\def\bp{{\bf p}}
\def\bq{{\bf q}}
\def\br{{\bf r}}
\def\bx{{\bf x}}
\def\by{{\bf y}}
\def\bR{{\bf R}}
\def\bV{{\bf V}}

\def\R{{\mathbb R}}
\def\C{{\mathbb C}}
\def\N{{\mathbb N}}
\def\Z{{\mathbb Z}}
\def\one{\mbox{1 \kern-.59em {\rm l}}}

\def\msu{\mathfrak{s}\mathfrak{u}}
\def\mmu{\mathfrak{u}}

\def\reps{{representations }}
\newcommand{\trr}{\triangleright}
\newcommand{\trl}{\triangleleft}

\def\bit{\begin{itemize}}
\def\eit{\end{itemize}}

\def\({\left(}
\def\){\right)}
\def\diag{\mbox{diag}}
\def\refeq#1{(\ref{#1})}
\def\gcheck{g^{\vee}}

\def\tens{\otimes}
\def\Pint{\int \!\!\!\!\!\!P}
\def\rep{representation }

\def\d{\delta}\def\D{\Delta}\def\ddt{\dot\delta}

\def\pa{\partial} \def\del{\partial}
\def\xx{\times}
\def\uno{\mbox{1 \kern-.59em {\rm l}}}  

\def\trp{^{\top}}
\def\inv{^{-1}}
\def\dag{{^{\dagger}}}
\def\pr{^{\prime}}

\def\rar{\rightarrow}
\def\lar{\leftarrow}
\def\lrar{\leftrightarrow}

\def\one{1\!\!1\,\,}
\def\im{\imath}
\def\jm{\jmath}

\newcommand{\tr}{\mbox{tr}}
\newcommand{\slsh}[1]{/ \!\!\!\! #1}

\def\vac{|0\rangle}
\def\lvac{\langle 0|}

\def\hlf{\frac{1}{2}}
\def\thlf{\tfrac{1}{2}}

\def\ove#1{\frac{1}{#1}}

\def\Box{\square}
\def\ZZ{\mathbb{Z}}
\def\CC#1{({\bf #1})}
\def\bcomment#1{}
\def\bfhat#1{{\bf \hat{#1}}}
\def\VEV#1{\left\langle #1\right\rangle}

\newcommand{\ex}[1]{{\rm e}^{#1}} \def\ii{{\rm i}}

\newtheorem{theorem}{Theorem}
\newtheorem{acknowledgement}{Acknowledgement}
\newtheorem{algorithm}{Algorithm}
\newtheorem{axiom}{Axiom}
\newtheorem{case}{Case}
\newtheorem{claim}{Claim}
\newtheorem{conclusion}{Conclusion}
\newtheorem{condition}{Condition}
\newtheorem{conjecture}{Conjecture}
\newtheorem{corollary}[theorem]{Corollary}
\newtheorem{criterion}{Criterion}
\newtheorem{definition}[theorem]{Definition}
\newtheorem{example}[theorem]{Example}
\newtheorem{exercise}{Exercise}
\newtheorem{lemma}[theorem]{Lemma}
\newtheorem{notation}[theorem]{Notation}
\newtheorem{problem}[theorem]{Problem}
\newtheorem{proposition}[theorem]{Proposition}
\newtheorem{remark}[theorem]{Remark}
\newtheorem{solution}{Solution}
\newtheorem{summary}{Summary}

\begin{document} 
\thispagestyle{empty}

\renewcommand{\title}[1]{\vspace{10mm}\noindent{\Large{\bf
#1}}\vspace{8mm}} \newcommand{\authors}[1]{\noindent{\large
#1}\vspace{5mm}} \newcommand{\address}[1]{{\itshape #1\vspace{2mm}}}


\begin{flushright}
UWThPh-2011-41
\end{flushright}

\begin{center}

\title{On Poisson geometries related to noncommutative \\[1ex] emergent gravity}

\vskip 3mm

\authors{Nikolaj Kuntner{\footnote{nikolaj.kuntner@dlr.de}} and Harold {Steinacker{\footnote{harold.steinacker@univie.ac.at}}}
}

\vskip 3mm

\address{ {\it Faculty of Physics, University of Vienna\\
 Boltzmanngasse 5, A-1090 Vienna (Austria)  }}

\vskip 1.4cm

\textbf{Abstract}

\end{center}

We study metric--compatible Poisson structures in the semi-classical limit of noncommutative emergent gravity.
Space-time is realized as quantized symplectic submanifold embedded in  $\R^D$, whose effective metric
depends on the embedding as well as on the Poisson structure.
We study solutions of the equations of motion for the
Poisson structure, focusing on a natural class of solutions 
such that the effective metric coincides with the embedding metric. This leads to 
$i$--(anti-) self-dual complexified Poisson structures in four space-time dimensions with 
Lorentzian signature. 
Solutions on manifolds with conformally flat metric are obtained and tools are developed which allow to  
systematically re-derive previous results, e.g. for the Schwarzschild metric. 
It turns out that the effective gauge coupling is related to the symplectic volume density, 
and may vary significantly over space-time. 
To avoid this problem, we consider in a second part space-time manifolds with compactified extra dimensions
and split noncommutativity,
where solutions with constant gauge coupling are obtained for several physically relevant geometries.


\newpage

\begin{spacing}{.3}
\thispagestyle{empty}

{
\noindent\rule\textwidth{.1pt}
 \tableofcontents 
\vspace{.4cm}
\noindent\rule\textwidth{.1pt}
}

\thispagestyle{empty}
\end{spacing}



\setcounter{page}1

\rhead{\thepage}
\cfoot{}

\section{Introduction}

In general relativity, space-time is described as a 3+1-dimensional manifold with pseudo-Riemannian
metric, governed by the Einstein equations. 
Despite the great success of  general relativity, there are 
good reasons why these classical notions of space and time should be questioned at very short distances. 
One idea is that space-time should not be viewed as classical manifold but 
as quantized or noncommutative (NC) space. This can be motivated by 
general arguments combining general relativity with quantum mechanics \cite{Doplicher:1994tu}, and also from 
string theory \cite{Amati:1988tn}.

A noncommutative space can be seen as 
quantization of a classical manifold with Poisson structure. The question then arises how such a 
Poisson structure is related to the (pseudo-) Riemannian metric, and how it affects other 
aspects of physics. These questions can be addressed in noncommutative emergent gravity,
where space-time is modeled by NC brane solutions of certain matrix models \cite{Steinacker:2010rh}. 
In the semi-classical limit, these branes are 
submanifolds $\cM^4 \subset \R^{D}$ with embedding metric $g_{\mu\nu}$ and a Poisson structure $\theta^{\mu\nu}$.
The effective metric is then given by $G^{\mu\nu} = e^{-\sigma} \theta^{\mu\mu'}\theta^{\nu\nu'} g_{\mu'\nu'}$,
where $e^{-\sigma}$ is a (dilaton-like) scalar field 
which determines the gauge coupling and measures the scale of noncommutativity.
The Poisson structure must satisfy\footnote{There are different approaches to noncommutative gravity which also involve a 
Poisson structure related to a metric, cf. \cite{madore}. Here we
focus on the matrix model framework.} certain equations of motions, which are similar to 
Maxwell equations on  curved manifolds. We will study these equations of motion for $\theta^{\mu\nu}$ 
on submanifolds with some given embedding geometry $g_{\mu\nu}$. 

In this paper, we  focus on the  simplest case where the effective metric
$G_{\mu\nu}$ coincides with the embedding metric $g_{\mu\nu}$. In the Euclidean case, there is an obvious class
of solutions for $\theta^{\mu\nu}$ with that property, given by (anti-) self-dual (A)SD Poisson structures.
It is easy to see that these are always solutions and in fact they are always minima 
of the action \cite{Blaschke:2010qj}.
In the case of Minkowski signature, the situation is less clear. 
While the effective metric $G$  has the same signature as $g$,  the causal structures are different provided the
Poisson structure $\theta^{\mu\nu}$ is real. 
In particular, there is no way that $G = g$. This is a priori not a problem since it is $G$ rather than $g$ which 
governs the physics on the brane, nevertheless it may seem a bit strange. 
This conclusion can be avoided by considering complexified Poisson structure as obtained by a 
Wick rotation 
$x^0 \to i t$. Then $G=g$ is indeed possible, and it holds for certain Poisson structures which 
are $i$-(anti-) self-dual $i$-(A)SD in the sense that $*_g \omega = \pm i \omega$ where $\omega$ 
is the symplectic structure associated 
with $\theta^{\mu\nu}$. This is the scenario under consideration in this paper. It remains to be seen whether 
such complexified or real Poisson structures are appropriate from a physical point of view. 

With this motivation, we study  $i$-(A)SD symplectic structures on physically 
relevant geometries. Such solutions have been obtained in \cite{Blaschke:2010ye} for the Schwarzschild geometry,
and also for certain Friedmann-Robertson-Walker geometries  \cite{Klammer:2009ku}. These solutions are largely determined 
by the asymptotics for $r \to \infty$; in particular for asymptotically flat spaces, 
 the Poisson tensor and $e^{-\sigma}$ should become constant, 
so that the physics approaches that of flat space. The reason is that 
 $e^{-\sigma}$ plays the role of a physical gauge coupling constant,
which is known to be constant to a very good approximation. 
However in the example of the Schwarzschild geometry, 
it turns out that the dilaton field $e^{-\sigma}$ has a non-trivial space-time dependence  
 near the horizon \cite{Blaschke:2010ye}. 
In fact there is a circle $S^1$ on the horizon where $\theta^{\mu\nu}$ becomes 
degenerate\footnote{cf. also \cite{taubes}.} and
$e^{-\sigma}$ goes to zero. Such a behavior is problematic from a physical point of view.

In the present paper, we achieve two things. First, we develop a more systematic understanding of
$i$-(A)SD Poisson structures on general 3+1-dimensional geometries, which allow to re-derive the 
above-mentioned results in a more systematic way. In particular, we show that  $i$-(A)SD Poisson structures
in general lead to a  foliation of space-time by two perpendicular 2--dimensional leaves. 
We also establish an appropriate version of the Darboux theorem in 4 dimensions, which provides
useful insights. Explicit $i$-(A)SD solutions are obtained for a class of spherically symmetric manifolds, and in particular for 
conformally flat metrics such as deSitter space. 
These results on $i$-(A)SD Poisson structures 
may also be of interest independent of the motivation considered here.

In the second part of this paper, we study a possible resolution of the  problems associated 
with non-constant $e^{-\sigma}$, by considering compactified extra dimensions 
$\cM^{2n} = \cM^{3,1}\times \cK$ where $\cK$ is some compact Riemannian space. 
Compactified extra dimensions are very well motivated
in physics, 
providing a link with particle physics via Kaluza-Klein compactification or intersecting branes.
It is important here that the entire space
$\cM^{2n}$ is symplectic, with symplectic structure relating the 
compact space $\cK$  with the non-compact space-time $\cM^{3,1}$. Such structures 
with ``split noncommutativity''  indeed arise as solutions of the IKKT matrix model 
\cite{Steinacker:2011wb}. They are  interesting for several
reasons, e.g. to the minimize the Lorentz violation due to the Poisson structure.
In section \ref{sec:extradim} we give explicit solutions for such symplectic structures on  physically
relevant space-times with compact extra dimensions. 
In particular, we show that for the Schwarzschild metric the overall $e^{-\sigma}$
can indeed be constant at least outside of the horizon. This solves the above-mentioned problem,
and opens up a promising approach towards
realistic physics within the framework of matrix models and emergent gravity.

Although we focus on complexified Poisson structures here, 
the  idea of Poisson structures relating space-time with compact extra dimensions is clearly more general,
and may  allow to obtain physically interesting solutions also for real Poisson 
structures. This should be studied elsewhere.

\section{Metric-compatible Poisson structures and (anti-) self-dual 2-forms}

We first explain the problem under consideration, which can be studied 
independent of its physical motivation given in the next section. 
Consider a 4-dimensional pseudo-Riemannian manifold $(\cM,g)$ with metric $g_{\mu\nu}$ with 
Minkowski signature,
carrying a Poisson structure $\{.,.\}$ encoded in a Poisson tensor field $\theta^{\mu\nu} = \{x^\mu,x^\nu\}$
in local coordinates. Assuming that it is non-degenerate, the inverse matrix $\theta^{-1}_{\mu\nu}$
defines a symplectic form 
\be
\o = \thlf \o_{\m\n}\dd x^\m \ww \dd x^\n, \qquad  \o_{\m\n} = \theta^{-1}_{\mu\nu} 
\ee
which is closed $\dd\o = 0$.
Our conventions are such that the spatial parts are positive, e.g. $\mbox{sign}(g)=(-,+,+,+)$. 
We will need the Hodge star operator $*_g$, which acts on 2-forms as
	$$*_g\o := \hlf\frac{1}{\sqrt{|g|}}\ g_{\m\a}\ g_{\n\b}\ \e^{\a\b\g\d} \left(\thlf\o_{\g\d}\right) \dd x^\m \ww \dd x^\n$$
with $*_g^2=-1$, whereas $*_g^2=+1$ in the Euclidean case. 
Here and in the following $|g|$ denotes $|\mbox{det}(g)|$.
We will use $\omega$ and $\theta^{-1}$ interchangeably in this paper, hoping that this does not cause confusion.

Given these structures, one can define another ``effective'' metric by
\be
G^{\m\n}:=e^{-\s}\th^{\m\a}\th^{\n\b}g_{\a\b} 
\label{eff-metric}
\ee
where 
\be
e^{-\s}:=\sqrt{\frac{\det\th^{-1}}{\det g}}.
\label{sigma-def}
\ee
This article studies symplectic (or Poisson) structures $\th$ which
are compatible with 
a given metric in the following sense 
\be
\label{Gisg}
G_{\m\n}=g_{\m\n} .
\ee
This condition becomes more transparent in terms of the following tensor
	\be
		\label{almost complex structure}
			\cJ^\m_\n := e^{-\frac{\s}{2}}\th^{\m\a}g_{\a\n}.
	\ee
It follows that $(\cJ^2)^\m_\n = -e^{-\s}(\th^{\m\a}\th^{\r\b}g_{\a\b})g_{\r\n}$ and we can write
	$$G^{\m\n}=-(\cJ^2)^\m_\r g^{\r\n}.$$
The condition $G_{\m\n} = g_{\m\n}$ is therefore equivalent to
	$$(\cJ^2)^\m_\n = -\d^\m_\n,$$
which is to say that $\cJ$ is an almost complex structure. In matrix notation this can be written as
	\begin{eqnarray*}
		\cJ&:=&e^{-\frac{\s}{2}}\ \th\  g, \qquad 
		\cJ^2=e^{-\s}\ (\th\  g\  \th)\ g \\
		G\inv&=&e^{-\s}\ \th\  g\  \th^T 
		= -\cJ^2\ g\inv 
\end{eqnarray*}
so that
\be
     G = g\ \ \Longleftrightarrow\ \ \cJ^2 = -\uno.
\ee
Therefore on a Riemannian manifold $\cM$, the metric compatibility condition $G=g$ amounts to 
the statement that $(\cM,\omega,\cJ)$ is  
almost-K\"ahler with almost-Kahler metric $e^{- \sigma/2} g_{\mu\nu}$  (but not with $g_{\mu\nu}$!). 
There is  considerable literature on the subject of four dimensional almost K\"ahler manifolds with given 
geometry, including notably the case of compact Einstein manifolds \cite{almost-k-einstein}.
However, the compatibility condition \eq{Gisg} requires finding almost-Kahler structures 
in the {\em conformal class} of the metric $g_{\mu\nu}$. 
Leaving aside global obstructions for the existence of symplectic structures (cf. \cite{MR2503498,taubes}),
it is easy to see that for 4-dimensional Riemannian manifolds, such compatible 
symplectic structures  
are in one-to-one correspondence with 
(non-degenerate) closed (A)SD 2-forms $\omega = \pm *_g\o$. 
For related work on this type of 
metric-compatible symplectic or Poisson manifolds see e.g. \cite{Blaschke:2010rg,Blaschke:2010qj,Arnlind:2011rz}.

However, the focus of this paper is the case of Lorentz signature, where the correct formulation of 
the problem is less obvious.
In that case, the compatibility condition $\cJ^2=-\uno$ 
requires to consider complexified symplectic structures, and a suitable reality condition must be imposed. 
We will first give an appropriate reality condition which allows to generalize the results of the 
Riemannian case to the case of Lorentz signature. Then 
the compatibility condition \eq{Gisg} is equivalent to the problem of finding 
closed complexified 2-forms which satisfy
\be
*_g\o = \pm i \o.
\ee
Such forms will be denoted as $i$-(A)SD,
and can be interpreted in terms of a complexified Poisson structure. 
We  give an appropriate normal form for such $i$-(A)SD structures 
which holds on a local neighborhood $\cU\subset \cM$, by establishing suitable Darboux-type coordinates. 
We then establish some results and tools towards finding explicitly such symplectic 
(or at least Poisson) structures
for a given metric $g_{\mu\nu}$. Of particular interest is the associated  function $e^{-\s}$,
which is related to the gauge coupling ``constant'' via \eq{YM-coupling}.
As such it must be non-vanishing and at least asymptotically constant 
on asymptotically flat space-times. 
This sets the boundary conditions for our problem, 
and justifies $\det \theta \neq 0$ on physical grounds.

In general, there may be global obstructions for such 
non-degenerate $\omega$, cf. \cite{MR2503498,taubes}. This
can be seen explicitly e.g. for the case of the Schwarzschild metric \cite{Blaschke:2010ye}.
To avoid this problem, we propose in the second part of this paper to consider higher-dimensional manifolds 
with a product (or a fiber bundle) structure $\cM^{2n} = \cM^{3,1}\times \cK$, and study symplectic 
structures such that $G=g$ and $e^{\sigma} = const$. 
This is motivated by the physical requirement that the gauge coupling constant should be 
constant to a very good approximation. 
This may provide an appropriate way to relax the almost-Kahler condition in 4 dimensions,
thus considerably extending the class of available geometries.

\section{Physical background: matrix models and emergent NC gravity}
\label{sec: motivation}

The problems under consideration here arise in the study of NC brane configurations in the 
following type of Yang-Mills matrix model
	\be
	\label{matrix model action}
	S=-(2\p)^n\ \mbox{Tr}\left( \tfrac{1}{4}\ [X^a,X^b] [X^{a'},X^{b'}] \eta_{aa'} \eta_{bb'} \right).
	\ee
Such models arise both in the context of string theory \cite{Ishibashi:1996xs} and NC gauge theory \cite{Douglas:2001ba}.
The $X^a$ are abstract infinite dimensional hermitian matrices or operators acting on a 
separable Hilbert space, with 
an index $a$ running from $1$ to $D$. The unphysical metric $\eta_{ab} = \diag(-1,1,...,1)$ fixes the signature of the theory.
The action has the following gauge symmetry
	$$X^\m \rar U X^\m U^{-1},$$
where $U$ is a unitary operator resp. matrix, and the equations of motion are given by 
$$[X^a,[X^b,X^{a'}]]\eta_{aa'}=0.$$
A simple example of a solution is given by the Moyal-Weyl quantum plane $\R^4_\theta$,
\begin{align}
 [X^\m,X^\n] &=\bar{\th}^{\m\n},\qquad \m=1,...,4,  \nn\\  
  X^i &=0,   \ \ \ \qquad  i=1,...,D-4, 
\end{align}
where $\bar{\th}^{\m\n}$ is a constant anti-symmetric matrix. 
This can be interpreted as canonically quantized 4-dimensional Minkowski space embedded in $\R^{D}$.
The matrices $X^\mu$ then generate the (quantized) algebra $\cA \cong End(\cH)$ of functions on $\R^4_\theta$. 
We will also consider solution or configurations corresponding 
to more general embeddings of $2n$--dimensional submanifolds (``branes'') in $\R^{D}$.
In all these configurations, the matrices
 $X^a$ are  interpreted as quantized embedding maps of some Poisson manifold embedded in $\R^D$,
\be
X^a \sim x^a: \quad \cM \hookrightarrow \R^D .
\ee
Here $\sim$ denotes the semi-classical limit as explained below.
Such configurations can be interpreted as quantized Poisson manifolds $(\cM,\theta^{\mu\nu})$
embedded in $\R^D$. 
The matrix model action (\ref{matrix model action}) then governs the dynamics of the 
quantized space-times $\cM$.


\paragraph{The semi-classical limit.}

The effective geometry and basic physical aspects of the resulting noncommutative emergent gravity model 
can be understood in the semi-classical limit, where commutators are replaced by Poisson brackets.
To understand this, consider a  Poisson or symplectic manifolds $(\cM,\theta^{\mu\nu})$,
together with a quantization map 
\be
\begin{array}{rcl}
\cI: \quad \cC(\cM) &\to& \cA\,\,\subset \,\, End(\cH)\, \\
 f(x) &\mapsto& \hat f
\end{array}
\label{quant-map} 
\ee
which depends on the Poisson structure $\theta^{\mu\nu}$, 
and satisfies\footnote{The precise definition of this
limiting process is non-trivial and there
are various definitions and approaches. 
Here we simply
assume that the limit and the expansion in $\theta$ exist
in some appropriate sense.}
\be
\cI(f g) - \cI(f)\cI(g) \,\, \to \,\, 0 \quad\mbox{and}\quad
\frac 1\theta \Big(\cI(i\{f,g\}) - [\cI(f),\cI(g)]\Big) \,\, \to \,\, 0 
\qquad \mbox{as}\quad \theta \to 0 .
\label{poisson-comp}
\ee
Here $\cC(\cM)$ denotes some space of functions on $\cM$, 
and $\cA$ is interpreted as quantized algebra of 
functions on $\cM$. 
One can then define a star product on $\cC(\cM)$ as follows
\be
f \star g := \ \cI^{-1}(\cI(f) \cI(g)) 
\ = \ f\cdot g+\tfrac{i}{2}\{f,g\}+O(\th^2) .
\label{star-product}
\ee 
Then the commutator behaves like the Poisson bracket up to first order in $\th$, i.e. 
$[f\ \overset{\star}{,}\ g]:=f\star g-g\star f=i\{f,g\}+O(\th^2)$. 
The ``semi-classical limit'' now amounts to replacing all commutators by Poisson brackets, 
and the noncommutative product by the classical product of functions.
\be
\com{\hat{f}}{\hat{g}} \sim i\{f,g\} = i\th^{\m\n}\pa_\m f \pa_\n g.
\ee
Here $\th^{\m\n} = \{x^\mu,x^\nu\}$ in some given coordinate system. 
Star products are useful because they allow a systematic expansion of the noncommutative 
structure in powers of $\theta$, within the framework of classical geometry.

\paragraph{Effective geometry.}

Now consider a  $2n$--dimensional Poisson manifold  embedded in $\R^D$,  via Cartesian 
embedding functions $x^a: \cM \hookrightarrow \R^D$. We can then consider its quantization 
as in \eq{quant-map}, which provides in particular quantized embedding functions $X^a = \hat x^a$
that generate $\cA \subset End(\cH)$.
This constitutes the class of backgrounds in the matrix model of interest here. 
One can then establish the basic geometric and physical properties of such backgrounds in the matrix model
in the semi-classical limit of the theory
\cite{Steinacker:2010rh},\cite{Blaschke:2010rg},\cite{Blaschke:2010qj}.
The  point is that both the Poisson structure $\th^{\m\n}$ 
as well as the embedding of $\cM$ are not arbitrary, but should be solutions of
the equations of motion governed by the action of our model. Here we focus on their semi-classical limit.
Denote the  \emph{induced metric} on $\cM \subset \R^D$ as
	$$g_{\m\n}:=  \del_\mu x^a \del_\nu x^b \eta_{ab} . $$ 
Then the \emph{effective metric} turns out to be
$G^{\m\n}=e^{-\s}\th^{\m\a}\th^{\n\b}g_{\a\b}$ as given in \eq{eff-metric}, with
	\be
	\label{ehochs}	
		e^{-(n-1)\s}:=\sqrt{\frac{\det\th^{-1}}{\det g}} .
	\ee
The positive quantity $e^{-\s}$ 
is of crucial importance \cite{Steinacker:2010rh} as explained below: it determines the gauge coupling. 
Thus for flat space-time, far away from any perturbations, it must clearly be constant and non-vanishing.  
This provides an important guideline in our search for solutions $\th^{\m\n}$ of the theory.

The origin of the effective metric $G^{\m\n}$ can be understood by 
considering the semi-classical limit of the kinetic term of a scalar field $\Psi$ on $\cM$:
\begin{align}
	S[\Psi]&=-(2\p)^n\Tr [X^a,\Psi] [X^b,\Psi]\eta_{ab} \nn \\
   & \sim \int \dd^nx\ \mbox{Pf}(\theta^{-1})\, (\th^{\m\n}\pa_\m x^a\pa_\n\Psi)(\th^{\r\s}\pa_\r x^b\pa_\s\Psi)\eta_{ab} \nn\\
&=\int \dd^nx\ |G^{\m\n}|^\hlf G^{\n\s}\pa_\n\Psi\pa_\s\Psi
\,.
\label{scalar-action}
\end{align}
Here the integral measure is given by the symplectic volume $\mbox{Pf}(\theta^{-1}) \sim \sqrt{\det\th^{-1}}$, 
cf. \cite{Steinacker:2010rh}.
Using the effective metric and the associated covariant derivative, 
one can show that the equations of motion  in the semi-classical limit imply 
(see e.g. (99)  in \cite{Steinacker:2008ya})
\be
\label{eom}
\nabla_\m(e^{\s} G^{\m\m'}\th\inv_{\m'\n'} G^{\n\n'})=  e^{-\s}\th^{\n\r}\pa_\r \h,
\qquad \h:=\tfrac{1}{4}\, e^\s\ G^{\m\n}g_{\m\n}.
\ee
Here  the covariant derivative $\nabla$ is always taken with respect to $G$. 
This equation is reminiscent of the Maxwell equations for the electromagnetic
field tensor on a general background. 
On a four-dimensional manifold $\cM$ with $G_{\m\n} = g_{\m\n}$ it follows that
	$\h= e^\s$,
so that the matrix model equations of motion (\ref{eom}) reduce to 
	\be
	\label{D Theta}
	 \nabla^\m\th\inv_{\m\n} =0,
	\ee
which is equivalent to 
	\be
	\label{co-closedness}
	 \dd *_g \th\inv = \dd *_G \th\inv = 0.
	\ee
We will see that any closed form $\th\inv$ with $G=g$ satisfies a Hodge self-duality relation, 
so that the equations of motion are satisfied identically.
Using a different argument, 
we will show in section \ref{sec:extradim} that  configurations with 
$G=g$ and $e^\sigma = const$ are also always solutions of (\ref{eom}) in the higher-dimensional case.

\paragraph{Noncommutative gauge theory.}

To explain the  significance of the function $e^{\sigma}$, we  briefly discuss 
gauge fields in the matrix model. Gauge fields arise as fluctuations of the matrices $X^a$
around a {\em stack} of coinciding NC brane configurations. 
More specifically, if $\bar X^a$ realizes a NC brane $\cM \subset \R^D$ as discussed above, 
then the following block--matrix configuration
\be
X^a = \bar X^a \otimes \one_n 
\label{coinciding-branes}
\ee
is interpreted as stack of coinciding NC branes. The point is that fluctuations around such a background 
behave as $\msu(n)$--valued gauge fields coupled to the effective metric $G_{\mu\nu}$,
with effective action \cite{Steinacker:2010rh}
\bea
S_{YM}[\cA] &\sim&
\frac{1}{4}\int d^{2n} x\ e^{\sigma}\sqrt{|G_{\mu\nu}|}\ G^{\mu\mu'} G^{\nu\nu'} 
\tr(F_{\mu\nu}\,F_{\mu'\nu'})\,\, 
+\,\, S_{NC} .
\label{action-expanded-2}
\eea
Here $S_{NC}$ is additional term which for 4-dimensional branes reduces to $\frac 12 \int \eta F\wedge F$.
Therefore the effective gauge coupling ``constant'' is given by
\be
g^2  \sim e^{-\sigma} .
\label{YM-coupling}
\ee
On $\R^4_\theta$, \eq{action-expanded-2} can be seen very easily 
by expanding
$$
X^\m=\bar X^\m-\th^{\m\n}A_\n.
$$
Then $A_\mu$ indeed transforms like a $\mmu(n)$-valued gauge potential
$$
A_\m \rar U  A_\m U^{-1}+i\ U \pa_\m U^{-1},
$$
 and the field strength $F_{\m'\n'}= \pa_\m A_\n -  \pa_\n A_\m + i[A_\m,A_\n]$ is encoded in the matrix commutator,
$$
\,[X^\mu,X^\nu] = -i \bar\theta^{\mu\mu'} \bar\theta^{\nu\nu'}
(\bar\theta^{-1}_{\mu'\nu'} + F_{\mu'\nu'}).
$$ 
However, the trace-$U(1)$ components of $A_\mu$ should  be interpreted
as fluctuation of the embedding  of the brane, which is part of the effective metric
$G_{\m\n}$ on $\cM \subset \R^D$. Then the derivation of \eq{action-expanded-2}
becomes somewhat more technical, see \cite{Steinacker:2008ya,Steinacker:2007dq}.

The fact that the gauge coupling is not a constant but 
a field is not surprising in view of string theory, where $e^\sigma$ plays the role of the dilaton. 
However, this means that $e^{\sigma}$ should be constant to a very good approximation,
in order to be compatible with stringent experimental bounds on the variation of the fundamental coupling 
constants.

\section{Metric-compatible Poisson structures and (anti-)self-duality}

In this section,
we establish the precise relation between the requirement of metric compatibility 
$g = G$ and generalized (anti-) self-duality in 
4 dimensions, and give a point-wise normal form for such $\theta^{\mu\nu}$.
While this reduces to basic linear algebra in the Euclidean case,  the Minkowski case 
is more tricky, because it requires a suitable complexification of the 
symplectic structure. We discuss this case in detail, and establish a number of useful  formulae 
along the way. This provides the basis for the extension to a local neighborhood $\cU\subset \cM$
in  section 5.

Consider a four-dimensional (pseudo-)Riemannian manifold with Euclidean or Lorentzian signature. 
Then the relation $G_{\m\n} = g_{\m\n}$  
can be written using \eq{sigma-def} as
	$$\th^{\m\a}\th^{\n\b}g_{\a\b} = \sqrt{\det\th\,\det g}\ g^{\m\n},$$
which can easily be solved locally. 
For any $p\in\cM$ we can choose a basis  such that the metric takes diagonal form 
$g_{\a\b} = \diag(g_{00},g_{11},g_{22},g_{33})$ with $g_{ii}>0$ for $i=1,2,3$, and $\e = \pm 1$. 
By considering  the diagonal entries of this relation, 
it follows easily that
the most general Poisson structure which satisfies $G=g$ is given by
	\be
	\label{G=g solution}	
\th^{\m\n} =
	\frac 1{\sqrt{\det (g)}}\,\begin{pmatrix}
		0 & -\e\ \sqrt{g_{22}g_{33}}\ f_3 & -\e\ \sqrt{g_{11}g_{33}}\ f_2 & -\e\ \sqrt{g_{11}g_{22}}\  f_1 \\ 
  		\e\ \sqrt{g_{22}g_{33}}\ f_3 & 0 & \ \ \ -\sqrt{g_{00}g_{33}}\ f_1 & \ \ \ \ \ \sqrt{g_{00}g_{22}}\ f_2 \\
   		\e\ \sqrt{g_{11}g_{33}}\ f_2 &  \ \ \ \sqrt{g_{00}g_{33}}\ f_1 & 0 & \ \ \ -\sqrt{g_{00}g_{11}}\ f_3 \\
    		\e\ \sqrt{g_{11}g_{22}}\ f_1 & -\sqrt{g_{00}g_{22}}\ f_2 & \ \ \ \ \sqrt{g_{00}g_{11}}\ f_3 & 0 \\
	\end{pmatrix} .
	\ee
Since we  assume $g_{ii}>0$ for $i=1,2,3$, 
this solution has the remarkable feature that\footnote{we will omit the index of the Hodge star operator from now on,
as the induces metric $g$ coinces with the embedding metric $G$.} 
$$*\ \th\inv = \e\ \sqrt{\mbox{sign}(g_{00})}\ \th\inv.$$
This means that $\th\inv$ is ($i$-)(A)SD, depending on $\e$ and on the signature of the induced 
metric.
In order to define a symplectic structure, the three functions $ f_1$, $ f_2$ and $ f_3$ have to be chosen in such a way
that $\th\inv$ is closed.  Then 
	\begin{align*}
		*\ \th\inv &= \pm\ \th\inv\ \ \hspace{1cm} \mbox{on\ a\ Euclidean\ metric,}\\
		*\ \th\inv &= \pm\ i\ \th\inv \hspace{1cm} \mbox{on\ a\ Lorentzian\ metric}
	\end{align*}
implies that $\th\inv$ is also co-closed. 
The corresponding coupling is then given by $e^{-\s}=(\sum_{i=1}^3 f_i^2)\inv$ 
in the representation \eq{G=g solution}.
Together with (\ref{co-closedness}),
this implies the following result:

\begin{lemma}
On four-dimensional (pseudo-)Riemannian manifold with Euclidean (or Lorentzian) signature, 
any symplectic form satisfying $G=g$ is ($i$-) (A)SD, and solves the 
equations of motion \eq{D Theta} of matrix model. 
\end{lemma}

Conversely, we will show that any ($i$-)(A)SD symplectic form -- subject to a certain reality condition 
in the case of Lorentzian signature -- satisfies $G=g$ and is hence a solution of the e.o.m. \eq{D Theta}, 
with local form given by \eq{G=g solution}.
To understand this, we  consider the Euclidean and Minkowski case separately.

\subsection{Euclidean signature}

It is clear that any symplectic structure $\theta^{-1}$ which is (A)SD with respect to the induced metric $g$
 satisfies $d *_g \th\inv = 0$, since $\th\inv$ is closed by assumption.
Moreover, it turns out that all such forms automatically also satisfy $G=g$ 
in the Euclidean case, so that the matrix model equations of motion (\ref{eom}) are satisfied.
This is best understood by bringing $\theta^{\m\n}$ 
into {\em normal form} at any given $p\in\cM$ as follows:
we can always find local coordinates such that 
$g=\mbox{diag}(1,1,1,1)$ at $p\in \cM$, and using an appropriate $SO(4)$ rotation the antisymmetric tensor $\th$ takes the form
\be
\th^{\m\n}=\mbox{det}(\th)^{\frac{1}{4}}
			\begin{pmatrix}
				0 & {-\e\ \a} & 0 & 0 \\
  				\e\ \a & 0 & 0 & 0 \\
   				0 & 0 & 0 & -\a\inv \\
    				0 & 0 & \a\inv & 0
			\end{pmatrix}
\label{theta-standard-euclid}
\ee
at  $p\in \cM$ with $\e=\pm1$. Therefore the corresponding symplectic form is
	\be
		\label{Euclidean Theta inv}
		\th\inv = \mbox{det}(\th\inv)^{\frac{1}{4}} \left( \e\ \a\inv\ \dd x^0 \ww \dd x^1 +  \a\ \dd x^2 \ww \dd x^3 \right)
	\ee
with dual 
	\begin{eqnarray*}
	*\ \th\inv &=& \e\ \mbox{det}(\th\inv)^{\frac{1}{4}} \left( \e\ \a\ \dd x^0 \ww \dd x^1 + \a\inv\ \dd x^2 \ww \dd x^3 \right).
	\end{eqnarray*}
Note that $\th\inv$ is  (A)SD if and only if $\a=\a\inv$, in which case
$\e$ indicates whether $\th\inv$ is self-dual or anti-self-dual. 
The effective metric is given in matrix notation by
	$$	G\inv
		:= e^{-\s} \th\ g\inv \th^T \\
		= -\sqrt{\det\th\inv}\ \th^2 \\
		= \mbox{diag}(\a^2,\a^2,\a^{-2},\a^{-2}).
	$$
This yields \cite{Steinacker:2010rh}
	$$*\ \th\inv=\e\ \th\inv
	\hspace{.3 cm} \Leftrightarrow \hspace{.3 cm}
	\a^2=1
	\hspace{.3 cm} \Leftrightarrow \hspace{.3 cm}
	G=\mbox{diag}(1,1,1,1)=g .$$ 
In particular, we can locally write an (A)SD symplectic form as
	\begin{eqnarray*}
		\th\inv
		&=& \mbox{det}(\th)^{-\frac{1}{4}}  (1+\e\ *)\ \dd x^2\ww\dd x^3.
	\end{eqnarray*}	
This applies in particular to the  Euclidean  Moyal-Weyl quantum plane $\R^4_\theta$ with $\theta^{\mu\nu}=\mbox{const}$,
where  $e^{-\s}=\mbox{det}(\th)^{-\frac{1}{2}}$ is clearly constant.

\subsection{Lorentzian signature: $i$-(anti-) self-duality and normal form}

We would like to establish the converse of Lemma 1 also in the Lorentzian case.
Even though this is still a point-wise question,
the argument of the previous section
does not generalize in a straightforward way, because we need to consider complexified symplectic forms.
In particular, we must identify a suitable reality condition which appropriately reduces the degrees of freedom. 
To understand the issue, assume that $g=\mbox{diag}(-1,1,1,1)$ at $p\in \cM$. 
For real Poisson tensors, the same local ansatz (\ref{Euclidean Theta inv}) leads to 
	$$G^{-1}=e^{-\s}\ \th\ g\ \th^T
	=\mbox{diag}(\a^2,-\a^2,\a^{-2},\a^{-2}).$$
This has the time component in the ``wrong'' slot, and there is no way to satisfy $G\sim g$. 
However, as  suggested by the factors of $\sqrt{g_{00}}$ in the general solution (\ref{G=g solution}), considering a purely imaginary component $\theta^{0i} = \{x^0,x^1\} \in i\, \R$ will automatically adjust this ``sign error''. 
This can be interpreted in terms of a  Wick rotation $x^0 \to i\, t$, resp. $X^0\to i\,T$ for matrices. 
The basic example is given by the Moyal solution in Minkowski space
\be
\th\inv_{\m\n} = c
	\begin{pmatrix}
		0 & \e\ i & 0 & 0 \\ 
  		-\e\ i & 0 & 0 & 0 \\
   		0 &  0 & 0 & 1 \\
    		0 & 0 & -1 & 0 \\
	\end{pmatrix}
\label{Moyal-Minkowski-SD}
\ee
with $c\in \R$, which gives indeed $G^{-1} = \diag(-1,1,1,1)$. We want to study more generally closed $i$-(anti-) self-dual 2-forms on 
manifolds with Lorentzian signature satisfying $G=g$. 
However,  in contrast to the Euclidean case not every $i$-(anti-) self-dual form satisfies $G=g$,
since multiplication by a constant phase
$\th\inv \to e^{i\a}\th\inv$ leads to  $G \to e^{-2i\a}G$.
We therefore impose the following  
reality condition for the Poisson tensor\footnote{
  It might be tempting to impose the following reality condition
  $(\cJ^\mu_\nu)^* = - g^{\mu\mu'}g_{\nu\nu'} \cJ^{\nu'}_{\mu'}$. However, this is not 
  compatible with Lorentz boosts. }
on a submanifold 
$\cM \subset \R^D$ with Minkowski signature
\be
\mbox{Pf}(\theta^{\mu\nu}) \in  i\,\R .
\label{reality-minkowski}
\ee
As a consequence, $\frac{\det g}{\det \theta^{-1}}$ and  $e^{-\s}$ 
are real and positive both in the Euclidean and in the Minkowski case, and analyticity in the $X^a$ can be formally preserved.
Note that then the  symplectic density $\mbox{Pf}(\theta^{-1})$ also 
yields the appropriate factor $i$ in front of the action such as \eq{scalar-action} in the Minkowski case.

To proceed, the following  observation will be crucial:
any $i$-(A)SD form can be written as\footnote{recall that 
the $\e = \pm 1$ indicates the SD resp. ASD case.}
\be
\th\inv=(1-\e\ i\ *)F, \qquad \mbox{where}\ \ F:= \mbox{Re}(\th\inv) \ \ \mbox{is {\em real}} .
\label{theta-F-rep}
\ee
This is a stronger statement than the obvious decomposition
$\o=\thlf(1-i\ *)\ \o+\thlf(1+i\ *)\ \o$ due to $*^2 = -1$, because $F$ is real.
To see this, note that 
\begin{align}
 \th\inv   
	&= \thlf(1-\e\ i\ *)(\mbox{Re}(\th\inv) +i\  \mbox{Im}(\th\inv)) \nn\\
	&= \thlf(\mbox{Re}(\th\inv) +\e * \mbox{Im}(\th\inv)) + i\ \thlf(-\e *\mbox{Re}(\th\inv) +\mbox{Im}(\th\inv))
\end{align}
which implies 
\be
 \mbox{Im}(\th\inv) = -\e *  \mbox{Re}(\th\inv) ,
\ee
and \eq{theta-F-rep} follows. 
In the case of the Moyal-Weyl solution \eq{Moyal-Minkowski-SD}, one finds
\be
	\label{Minkowski Moyal inv}
		\th\inv=c\ (1-\e\ i\ *)\ \dd x^2\ww\dd x^3, \qquad c \in \R
\ee
corresponding to $F =c\ \dd x^2\ww\dd x^3$. 

We can now compute $e^{-\sigma}$ in terms of this $F$.
Recall that the Pfaffian of a (skew-symmetric) $2$-form $\o$ is defined as
	$$ \mbox{Pf}(\o)
	:= \tfrac{1}{8} \e^{\a\b\g\d} \o_{\a\b} \o_{\g\d} 
	= \tfrac{1}{4} \sqrt{|\det g|}\ (* \o)_{\m\n} \o^{\m\n} ,$$
and satisfies $\mbox{Pf}(\o)^2=\mbox{det}(\o)$. For the 2-form $\th\inv = (1-\e\ i\ *)F$ this gives
	\begin{eqnarray}
	\label{Pfaffian}
		\mbox{Pf}(\th\inv) 
		= \tfrac{1}{2} \sqrt{|\det g|} \left((* F)_{\m\n} F^{\m\n}+\e\ i\ F_{\m\n} F^{\m\n} \right). 
	\end{eqnarray}
In particular, 
 \begin{eqnarray}
		e^{-\s} = \frac{\mbox{Pf}(\th\inv)}{\sqrt{\det(g)}} = \pm\thlf F_{\m\n} F^{\m\n}
	\end{eqnarray}
since $\det(g) < 0$. 
Now we can easily show the following result:
 \begin{lemma}
Let $\th\inv$ be an  $i$-(A)SD symplectic form. Then the reality condition  \eq{reality-minkowski}
holds if and only if the determinant of $F:= \mbox{Re}(\th\inv)$ vanishes. 
In that case, the effective metric  $G^{\m\n} := e^{-\s}\th^{\m\m'}\th^{\n\n'}g_{\m'\n'}$ 
satisfies $G_{\m\n} = \pm g_{\m\n}$, and there exists a Lorentz 
transformation for any given point on $\cM^4$ 
such that $\theta^{\mu\nu}$ has the form \eq{Moyal-Minkowski-SD} resp. 
(\ref{Minkowski Moyal inv}).
\end{lemma}
\textbf{Proof:}\\
Let $\th\inv$ be an $i$-(A)SD symplectic form. Then \eq{Pfaffian}
implies that the reality condition  \eq{reality-minkowski} is equivalent to 
\be
\mbox{Pf}(\theta^{\mu\nu}) \in  i\,\R 
\quad \Longleftrightarrow\quad (* F)_{\m\n} F^{\m\n} = 0 \quad \Longleftrightarrow \quad \mbox{det}(F)  = 0 .
\label{reality-minkowski}
\ee
This implies that $F$ has rank 2, and it is easy to see (see e.g. \cite{Sexl:1976pg})
that it can be brought into the form $F =c\ \dd x^2\ww\dd x^3$
or its Hodge dual using a suitable Lorentz transformation.
Then $G = \pm g$ follows, which
completes the proof. \hfill $\Box$

Together with Lemma 1, this characterizes the solutions of $G_{\m\n} = \pm g_{\m\n}$.
As a remark, notice that due to $\mbox{det}(F) \propto \mbox{det}(* F)$, 
the lemma can equally be formulated in terms of the imaginary part or $\th\inv$.
A multiplication of $\th$ with $i$ amounts to replacing $F$ with $*F$,
replacing $G$ by $-G$.


\section{Local normal form on four-dimensional Lorentzian manifolds}


In this section, we will investigate general properties of the metric-compatible symplectic form of interest here,
and establish a normal form on a local neighborhood $\cU\subset \cM$
in terms of certain adapted Darboux coordinates.
The main results are summarized as follows:
\begin{proposition}
Let $\th\inv$ be a symplectic form on a four-dimensional manifold with Lorentzian metric satisfying
the compatibility condition
	$$G^{\m\n}:=e^{-\s}\th^{\m\m'}\th^{\n\n'}g_{\m'\n'} =g^{\m\n} .$$
Then for every  $p\in\cM$ there is an open neighborhood  $p\in \cU\subset\cM$
and two real functions $\Phi,\Psi\in \cC^2(\cU)$ such that
	\be
	\label{thetaPhiPsi}	
		\th\inv = (1-\e\ i\ *_g)\ \dd\Phi\ww\dd\Psi 
	\ee
with 
	\be
	\label{neweom}	
		\dd(*_g\ \dd\Phi\ww\dd\Psi) = 0.
	\ee
The sign $\e=\pm1$ indicates whether $\th\inv$ is $i$-self-dual or $i$-anti-self-dual.
In local coordinates, the closedness condition reads
	\be
	\label{neweomcoord}	
		g^{\m\n}(\pa_\m\Phi\ \pa_\n(g^{\r\s}\pa_\s\Psi)-\pa_\m\Psi\ \pa_\n(g^{\r\s}\pa_\s\Phi))+g^{\r\s}(\D_g\Phi(\pa_\s\Psi)-\D_g\Psi(\pa_\s\Phi))=0,
	\ee
and we have
	$$e^{-\s} = (g^{\m\n}\pa_\m\Phi\pa_\n\Phi)(g^{\a\b}\pa_\a\Psi\pa_\b\Psi)-(g^{\m\n}\pa_\m\Phi\pa_\n\Psi)^2.$$
Furthermore, there is a local ``Darboux coordinate system'' $\phi^{D} = (\Phi,\Psi,\Phi',\Psi')$ on $\cU$, such that 
\be
\th\inv = \dd\Phi\ww\dd\Psi - \e\ i\ \dd\Phi'\ww\dd\Psi'.
\label{darboux-rep}
\ee
\end{proposition}
The proof is provided by the following two subsections.
Notice that this goes beyond the standard Darboux theorem, taking into account the metric compatibility
and the complex structure in the Lorentzian setting. The specific complexification
adds extra structure subject to \eq{reality-minkowski}. It turns out that 
the real and the imaginary part $\mbox{Re}(\th^{-1})$ and $\mbox{Im}(\th^{-1})$ of $\th^{-1}$
both satisfy the closedness separately, thus reducing the degrees of freedom of $\th^{-1}$.


\subsection{Implications of $\mbox{det}(F)=0$}
From
	\be
	\label{F conditions}	
		\mbox{det}(F) \propto (* F)_{\m\n} F^{\m\n}=0 \hspace{.5 cm} \mbox{and} \hspace{.5 cm} 
      e^{-\s}=\thlf F_{\m\n} F^{\m\n} \ne 0
	\ee
it follows that $F$ is a rank two tensor and can be written as $F=X\ww\ Y,$ 
where $X$ and $Y$ are two linearly independent 1-forms. In terms of components we have
	\be
	\label{FXY}	
		F_{\m\n} = g_{\m\m'}g_{\n\n'}F^{\m'\n'}
		= X_\m Y_\n - Y_\m X_\n,
	\ee
where  $X_\m=g_{\m\n}X^\n$ and $X^\m=g^{\m\n}X_\n$.
Since the relations (\ref{F conditions}) similarly hold for $* F$, we also have $* F=U \ww\ V$ 
with two linear independent 1-forms $U$ and $V$.

\begin{proposition}
Let $E^{XY}$ and $E^{UV}$ denote the distributions spanned by the vector fields $X,Y$ resp. $U,V$ defined above.
 These distributions are orthogonal complements with respect to $g$, i.e.
	$$T\cM=E^{XY}\oplus E^{UV},\ \ \ \mbox{with}\ \ \  E^{XY}\bot\ E^{UV}.$$
\end{proposition}
\textbf{Proof:}\\
First we observe that $* F$ can also be expressed in terms of $X$ and  $Y$:
	\begin{eqnarray*}
		(* F)_{\m\n}
		&=& \thlf\tfrac{1}{\sqrt{|g|}} g_{\m\a} g_{\n\b} \e^{\a\b\g\d} (X_\g Y_\d - Y_\g X_\d)\\
		&=& -\sqrt{|g|}\ \e_{\m\n\m'\n'} X^{\m'} Y^{\n'}.
	\end{eqnarray*}
Similarly,
$$F_{\m\n}=-(* (* F))_{\m\n}\propto\ \e_{\m\n\m'\n'} U^{\m'} V^{\n'}.$$
Therefore	
	$$\mbox{det}\left((U,V,X,Y)^T\right)
	= \e_{\a\b\g\d}U^\a V^\b X^\c Y^\d
	\propto F_{\a\b} F^{\a\b}
	\ne 0, $$
so that $X,Y,V,U$ span the entire tangent bundle. 
Now let $W$ be any vector field in $E^{UV}$. We have
$$(i_W F)_\a \propto \e_{\a\b\g\d}U^\b V^\g W^\d = 0 \hspace{.3 cm} \Longrightarrow \hspace{.3 cm} E^{UV} \subseteq \mbox{ker}(F).$$
On the other hand
	\begin{eqnarray*}
 		(i_W F)_\nu
		&=& \left(X_\m Y_\n - Y_\m X_\n \right) W^\m \\	
		&=& g(X,W)\ Y_\n - g(Y,W)\ X_\n.
	\end{eqnarray*}
Using the linear independence of $X$ and $Y$, it follows that
$$ \hspace{4 cm} g(X,W)=g(Y,W)=0 \hspace{0.5 cm} \forall\ W \in E^{UV}. \hspace{3.7 cm} \Box$$

\subsection{A version of the Darboux theorem for $i$-(anti-) self-dual 2-forms}

Consider again a closed $i$-(anti-) self-dual 2-form $\th\inv = (1-\e\ i\ *)F$ on $\cM$. Since $F$ is real, 
the closedness condition is equivalent to $F$ satisfying the Maxwell equations in vacuum:
	\be
	\label{Maxwell equations}	
		\dd \th\inv=0 \hspace{0.5 cm} \Longleftrightarrow \hspace{0.5 cm} \dd F=0, \hspace{0.5 cm} \dd* F = 0.
	\ee
Formulated in terms of a local vector potential $A$ with $F=\dd A$, these equations can be written as 
	$$\D_gA-(\dd*\dd *)A=0,$$
where $\D_g$ is the Laplace--de Rham operator, or
$\D_gA_\m-\nabla^\n\nabla_\m A_\n=0$ 
in a coordinate frame.
However due to the constraints (\ref{F conditions}), 
finding the general solution to the Maxwell equations in terms of the gauge potential on a curved space-time 
is difficult. We will therefore use the previously established form 
(\ref{FXY}) of $F$:
	\begin{eqnarray*}
		\dd* F
		&\hat=& \nabla_\n F^{\m\n} \ 
		= \tfrac{1}{\sqrt{|g|}}\ \pa_\n \left( \sqrt{|g|} F^{\m\n} \right)\\			
		&=& \tfrac{1}{\sqrt{|g|}}\ \pa_\n \left( \sqrt{|g|} (X^{\m}Y^{\n}-Y^{\m}X^{\n}) \right)\\	
		&=& Y^\n\pa_\n X^\m - X^\n\pa_\n Y^\m 
		+ \tfrac{1}{\sqrt{|g|}} \pa_\n (\sqrt{|g|}\  Y^\n) X^\m - \tfrac{1}{\sqrt{|g|}} \pa_\n (\sqrt{|g|}\ X^\n) Y^\m \\
		&=& \com{Y}{X}^\m + \mbox{div}(Y) X^\m + \mbox{div}(X) Y^\m 
	\end{eqnarray*}
where ``$\com{\cdot}{\cdot}$'' and ``$\mbox{div}$'' denote the Lie bracket resp. divergence of vector fields. 
Therefore for  $F$ of the form \eq{FXY}, the Maxwell equations $\dd* F = 0$ are equivalent to 
\be
\com{X}{Y} = \mbox{div}(Y) X + \mbox{div}(X) Y.
\ee
Similarly, from 
	$$\dd F=-\dd*(* F) \overset{!}{=} 0$$
we find the analogous equations, where $X$ and $Y$ are replaced by $U$ and $V$. We therefore established that
	\be
	\label{Integrability}	
		\com{X}{Y}\in \Gamma(E^{XY}), \hspace{.5 cm} \com{U}{V}\in \Gamma(E^{UV}),
	\ee	
where $\Gamma(E)$ are the vector fields over a distribution $E$ (i.e. the smooth sections of $E$). 
These are precisely the integrability conditions of the distributions $E^{XY}$ and $E^{UV}$ in terms of Lie brackets. 
This means that there exist 
two different but equivalent orthogonal foliations of the base manifold, such that 
	\be
	\label{TMEB}	
		T\cM = E^{XY}\oplus E^{UV}
		= TB^{XY}\oplus TB^{UV},
	\ee
where $B^{XY},B^{UV}$ are the leaves of the foliations. Furthermore, since we already established that the elements of $E^{UV}$ lie in the kernel of $F$, 
we have $${\cal{L}}_W F = \left( \dd i_W + i_W \dd \right)F = \dd (i_W F) + i_W (\dd  F) = 0,$$
where $W\in E^{UV}.$ By applying the Darboux theorem in the two dimensional submersion of $B^{XY}$, it follows
that the form $F$ (and consequently also $\th\inv$) can be written locally in terms of two real functions $\Phi,\Psi$
such that 
	$$F = \dd\Phi\ww\dd\Psi .$$
Analogous considerations lead to two further functions $\Phi',\Psi'$ such that $* F = \dd\Phi'\ww\dd\Psi'$. 
We therefore obtain a local ``Darboux coordinate system'' $\phi^{D} = (\Phi,\Psi,\Phi',\Psi')$ on $\cU\subset\cM$ 
indicated by a capital $D$, 
such that the $i$- (A)SD form $\th\inv$ takes the form
	\be
	\label{Darboux F}	
	\th\inv 	= (1-\e\ i\ *)F = \dd\Phi\ww\dd\Psi - \e\ i\ \dd\Phi'\ww\dd\Psi',
	\ee	
which establishes \eq{darboux-rep}.
Moreover using (\ref{TMEB}) it follows that in these Darboux coordinates $\phi^D$, the metric decomposes into two 
orthogonal blocks
	$$\hspace{.3 cm} g_{\phi^D}
	= \begin{pmatrix}
		g^{XY} & 0\\
  		0 & g^{UV}\\
	\end{pmatrix},
	\hspace{.3 cm} \mbox{with} \hspace{.3 cm} g^{XY}
	=		\begin{pmatrix}
				g(X,X) & g(X,Y)\\
  				g(X,Y) & g(Y,Y)
			\end{pmatrix},
	\hspace{.1 cm} g^{UV}
	=		\begin{pmatrix}
				g(U,V) & g(U,V)\\
  				g(U,V) & g(U,V)
			\end{pmatrix}$$
where $g^{XY}, g^{UV}$ denote the induced metrics. Here $X,Y$ and $U,V$ denote the two orthogonal 
components established earlier. 

This Darboux coordinate system $\phi^{D}$ is of course not unique. For example, the transformations 
	$$\Phi\mapsto f(\Phi),\ \Psi\mapsto \frac{\Psi}{f'(\Phi)},$$
where $f$ is any well behaved function, always leaves $F$ invariant:
	\begin{eqnarray*}
		F = \dd\Phi\ww\dd\Psi
		&\mapsto& \dd f(\Phi)\ww\dd \left(\Psi\frac{1}{f'(\Phi)} \right)\\
		&=& f'(\Phi)\dd \Phi \ww \left( \frac{1}{f'(\Phi)} \dd \Psi + \Psi \left(\frac{\pa}{\pa\Phi}\frac{1}{f'(\Phi)} \right) \dd \Phi \right)\\
		&=& \dd\Phi\ww\dd\Psi.
	\end{eqnarray*}
This transformation can be viewed as a symplectomorphism on $B^{XY}$.

\paragraph{Implications for $e^{-\s}$.}

Using these results, we can compute
	\begin{eqnarray*}
		e^{-\s}
		&=& \thlf\ F_{\m\n}F^{\m\n}\\
		&=& (g^{\m\n}\pa_\m\Phi\pa_\n\Phi)(g^{\a\b}\pa_\a\Psi\pa_\b\Psi)-(g^{\m\n}\pa_\m\Phi\pa_\n\Psi)^2\\
		&=& (\det g^{XY})\inv
		= -(\det g^{UV})\inv
		= \pm i (\det g_{\phi^D})^{-1/2}.
	\end{eqnarray*}
The relation $\det g^{XY} = -\det g^{UV}$ stems from the fact that $e^{-\s}$ changes sign under $F\mapsto * F$,
reflecting the different signatures.
The last equality $e^{-\s} = \pm i(\det g_{\phi^D})^{-1/2}$ 
can also be seen from the very definition (\ref{ehochs}) of $e^{-\s}$, since in Darboux coordinates (\ref{Darboux F})
 we have $\det \th\inv = \mbox{const}$. As a remark, notice that if $e^{-\s}$ is to be constant, then the blocks of the 
Darboux metric have to have constant determinant $\det g^{XY} = \det g^{UV}=\mbox{const}$.
	
\section{Explicit solutions for $\theta^{\mu\nu}$}

\subsection{A solution on conformally flat metrics}

As in the case of the Maxwell equations in vacuum, the metric enters into the equations of motion (\ref{neweom})
only via the Hodge star. Since the Hodge star acting on 2-forms in four dimensions is invariant under 
Weyl rescaling transformations 
$$g_{\mu\nu}\mapsto g'_{\mu\nu}=\frac{1}{f}\ g_{\mu\nu},$$
 it follows that each solution provides automatically also a solutions for 
the rescaled metric. 
This can also be seen directly from (\ref{G=g solution}).

The archetypal solution of the equations of motion with the form (\ref{thetaPhiPsi})
is given by the Moyal solution
	$$\bar\th\inv_{\m\n}:=c
		\begin{pmatrix}
			0 & {-\e\ i} & 0 & 0 \\
  			\e\ i & 0 & 0 & 0 \\
   			0 & 0 & 0 & 1 \\
    			0 & 0 & -1 & 0
		\end{pmatrix},
		\hspace{.5 cm}
		\bar\th\inv=c\ (1-\e\ i\ *)\ \dd x^2\ww\dd x^3,$$
with $c\in \R\backslash\{0\}$ and $e^{-\s}=c^2$.  Therefore $\bar\th\inv$ is also a solution for all 
conformally flat metrics $g_{\mu\nu}=\frac{1}{f}\ \h_{\mu\nu}$. In that case, we have
	$$e^{-\s}=\frac{\sqrt{\det \bar\th\inv}}{\sqrt{\det g}}= \frac{1}{\sqrt{|g_{\phi^D}|}} = c^2f^2,$$
which reduces to the Moyal case for $f=1$.
There are many examples of physically interesting spaces which admit a conformally flat metric,
see e.g.  \cite{Ibison:2007dv}. This includes Robertson-Walker space-times, in particular
de Sitter and anti-de Sitter space, with conformal factor given by $f=(1-||x||^2)^2$ and $f=(1+||x||^2)^2$, 
respectively. Another example which was studied from the present  point of view is given in  \cite{Klammer:2009ku}.

	
\subsection{Electrostatic solutions}

In the case of a diagonal metric $g=\mbox{diag}(g_{00},g_{11},g_{22},g_{33})$, we can write the equations (\ref{neweomcoord}) as
	\be
	\label{diageom}	
		\pa_\m \left( \sqrt{|g|}\ g^{(\r\r)}g^{\m\n}\left(\pa_{(\r)}\Phi \pa_{\n}\Psi-\pa_{\n}\Phi \pa_{(\r)}\Psi \right) \right)=0.
	\ee
Here the Einstein summation convention with respect to the free index $\r$ does not apply.
Furthermore, we now make an ``electrostatic ansatz''
	$$\pa_0\Phi=0,\hspace{.3 cm} \Psi=x^0.$$
Then the equations of motion reduce to
	\be
	\label{electrostatic eom}
		\pa_i \left( \sqrt{|g|}\ g^{00} g^{ij} \pa_j\Phi \right)=0
	\ee
where the latin index $i$ runs from $1$ to $3$, as well as the condition
	\be
	\label{metric condition}
		\pa_0 \left( \sqrt{|g|}\ g^{00}g^{ij} \right)=0.
	\ee
Assuming $\mbox{sign}(g)=(-,+,+,+)$ we have
	$$	e^{-\s}
		= (g^{\m\n}\pa_\m\Phi\pa_\n\Phi)(g^{\a\b}\pa_\a\Psi\pa_\b\Psi)-(g^{\m\n}\pa_\m\Phi\pa_\n\Psi)^2
		= -g^{00} \sum_{i=1}^3 g^{ij} (\pa_j\Phi)^2.
	$$
	

\subsubsection{Conformally flat metrics}

As a first example, we immediately re-establish that for any conformally flat metric $g_{\mu\nu}=\frac{1}{f} \h_{\mu\nu}$, 
the functions $\Psi=x^0$ and $\Phi=c\cdot x^1$ are a valid solution of (\ref{electrostatic eom}), 
since in this case $\sqrt{|g|}\ g^{00} g^{ij}=\d_{ij}=\mbox{const}$. We also obtain
	$$e^{-\s} = -g^{00} \sum_{i=1}^3 g^{ij} (\pa_j\Phi)^2 = c^2f^2.$$

\subsubsection{Radially symmetric metrics}
\label{The general case}
Our second example is another type of metric which is frequently encountered in physics, namely
	\be
	\label{SphericalCoordinates}
		\dd s^2 = -\k(r)\dd t^2+\frac{1}{\k(r)}\dd r^2+r^2 \left( \dd \vth+\sin^2(\vth)\dd \varphi^2 \right), \hspace{.3 cm} \k(r)>0. 
	\ee
This class of metrics is independent of $t$, so (\ref{metric condition}) is satisfied.
If we make the $\varphi$-independent separation ansatz 
	$$\Phi(r,\vth)=\phi(r) Y_l(\vth)$$
where $Y_l(\vth)$ are the spherical harmonics, then the equations (\ref{electrostatic eom}) reduce to an ordinary 
differential equation in $r$:
	\be
	\label{reom}
		\left( r\ \phi(r) \right)''-\frac{l(l+1)}{r\ \k(r)}\phi(r)=0.
	\ee
Furthermore
	\be
	\label{ehochsigma}
		e^{-\s} = -g^{00} \sum_{i=1}^3 g^{ij} (\pa_j\Phi)^2
		= \left( \phi'(r)\ Y_l(\vth) \right)^2  +  \frac{1}{r^2\ \k(r)} \left(\phi(r)\ Y_l'(\vth) \right)^2.
	\ee
\underline{The flat limit:}\\
For $\k(r)=1$ we recover the Minkowski metric in spherical coordinates. The differential equation is then just 
the well-known Laplace equation for the electric potential $\phi(r)$ in  electrostatics\footnote{If we change variables 
$\hat\phi(r):=r\ \phi(r)$, then the equation take the familiar form $\hat\phi''(r)-\frac{l(l+1)}{r^2}\hat\phi(r)=0.$}:
	$$\left( r\ \phi(r) \right)''-\frac{l(l+1)}{r}\phi(r)=0.$$
For $\underline{l=0}$, we are left with the condition $(r\ \phi(r))''=0$, which gives
\begin{align}
 \Phi(r)&=\frac{a}{r}+b,  \qquad
e^{-\s}   = \left( \frac{a}{r^2} \right)^2.
\end{align}
In fact, since for $l=0$ the differential equation (\ref{reom}) is independent of $\k(r)$, 
this radial point particle potential solution $\Phi(r)$ exists for all metrics of the type (\ref{SphericalCoordinates}).

For $\underline{l=1}$, we find 
	$$\Phi(r,\vth)=\left(\frac{a}{r^2}+b\ r\right)\cos(\vth).$$ 
For $a=0$, the potential $\Phi(r,\vth)=b\ r \cos(\vth)=b\ z$ reproduces the Moyal solution (\ref{Minkowski Moyal inv})
	\begin{eqnarray*}
		\bar\th\inv 
		&=& \pa_{\m}\Phi\ \pa_{\n}\Psi\ (1-\e\ i\ *)\ \dd x^\m\ww\dd x^\n \\
		&=& b \left( \e\ i\ \dd x\ww\dd y + \dd z\ww\dd t \right),
	\end{eqnarray*}
and 
$$
e^{-\s} =  b^2. 
$$
It should be noted that in the flat case, the solutions of (\ref{reom}) can be calculated for all $l$ in closed form, so that 
	\be
	\label{flat solution}
		\phi(r)=a\ \frac{1}{r^{l+1}}+b\ r^l.
	\ee
However, the relation
	$$\underset{r\rightarrow\pm\infty}{\mbox{lim}}e^{-\s}=\mbox{const}$$ 
only holds for the Moyal solution $l=1$. 

\underline{The general case:}\\	
Solving (\ref{reom}) for $\k(r)$ and plugging it into (\ref{ehochsigma}) one can show that with this ansatz, 
only flat space-time admits strictly constant $e^{-\s}$. Relaxing this requirement, 
we are interested in asymptotically flat $i$- (A)SD solutions which reduce to the Moyal case for $r\to\infty$. Hence we expect
	\begin{eqnarray*}
		g\rightarrow \h \hspace{.2cm} &\Longleftrightarrow&  \hspace{.2cm} \k(r) \rightarrow 1 \\
		&\downarrow& \\
		\th\inv \rightarrow \bar\th\inv \hspace{.2cm} &\Longleftrightarrow&  \hspace{.2cm} \phi(r) \rightarrow b\ r.
	\end{eqnarray*}
We can thus set $l=1$ and try to solve (\ref{reom}) for any $\k(r)$ of interest, together with $\phi(r) \rightarrow b\ r$ as 
boundary condition in the flat limit. Alternatively, since we were always considering a general metric of 
the type (\ref{SphericalCoordinates}) up to now, we can view (\ref{reom}) as equation in $\k(r)$ for given $\phi(r)$.
It is convenient to make the following ansatz  
	$$\phi(r)=b\ r\ \k(r)\ \varphi_\k(r) \hspace{.5 cm} \mbox{with} \hspace{.5 cm} \underset{\k\rightarrow 1}{\mbox{lim}}\ \varphi_\k(r)=1.$$
The resulting equation
	$$\left( r\ \k(r)\ \varphi_\k(r) \right)''-l(l+1)\varphi_\k(r)=0$$
can be solved for $\k(r)$ as follows
	$$\k(r)=\frac{1}{\varphi_\k(r)}\left(   \frac{l(l+1)}{r^2}  \int\limits_{1}^{r}\varphi_\k(\varrho)\left( r-\varrho \right)\dd \varrho +    \frac{c_1}{r} +  \frac{c_2}{r^2}  \right),$$
where $c_1,c_2$ are constants of integration, and $\varphi_\k(r)$ is any well-behaved function. 
This determines a metric of the type \eq{SphericalCoordinates} for any given $\varphi_\k(r)$.
For example, 
for $l=1$ and $r$-independent $\varphi_\k=1$, we obtain
	\begin{eqnarray*}
		\k(r)
		&=& \frac{2}{r^2}  \int\limits_{1}^{r}\left( r-\varrho \right)\dd \varrho  + \frac{c_1}{r} +  \frac{c_2}{r^2} \\
		&=& 1+ \frac{(c_1-2)}{r} + \frac{(c_2+1)}{r^2} \\
		&\equiv& 1- \frac{r_c}{r} + \frac{Q^2}{r^2}.
	\end{eqnarray*}
Therefore, the potential 
	$$\phi(r) = b\ r\ \k(r) = b \left( r - r_c +\frac{Q^2}{r} \right)$$ 
is a solution for the Reissner–Nordstr\"om metric. 
The corresponding tensor $\th^{\m\n}$ is not spherically symmetric, since one of the spatial directions is distinguished. 
Finally,
	\begin{eqnarray*}
		e^{-\s}
		&=& \left( \phi'(r)\ Y_l(\vth) \right)^2  +  \frac{1}{r^2\ \k(r)} \left(\phi(r)\ Y_l'(\vth) \right)^2 \\
		&=& \left(b\ r\ \k(r)\right)'^2\cos^2(\vth)  +  b^2\k(r) \sin^2(\vth)\\
		&=& b^2 \left(\left(1-\frac{Q^2}{r^2}\right)^2   +  \left(-\frac{r_c}{r}+\frac{Q^2}{r^2} \left(3-\frac{Q^2}{r^2}\right)\right)   \sin^2(\vth)\right). 
	\end{eqnarray*} 
For $Q=0$ we obtain the Schwarzschild metric with 
	$$e^{-\s} = b^2 \left(1-\frac{r_c}{r} \sin^2(\vth) \right)$$ 
and for $r_c=0$ we recover the flat limit $e^{-\s} = b^2$. This reproduces the result given in \cite{Blaschke:2010ye}. 
Notice that for $r=r_c$ there is a circle on the horizon where $e^{-\s}=0$. This means that 
$\th^{-1}$ is degenerate. We will propose a resolution of this issue later,
by considering compactified extra dimensions.


\section{Solutions on manifolds with extra dimensions}
\label{sec:extradim}
	
We have seen how to find non-degenerate closed 2-forms on four-dimensional Lorentzian manifolds which satisfy
	$$G^{\m\n}:=e^{-\s}\th^{\m\m'}\th^{\n\n'}g_{\m'\n'}
	=g^{\m\n}.$$
However, with the exception of the Moyal case, the invariant 
	$$e^{-\s}=\sqrt{\frac{\det \th\inv}{\det g}},$$
which acts as a coupling constant, always turned out to be space-time dependent. For example, on conformally flat metrics 
$g_{\mu\nu}=\frac{1}{f}\h_{\mu\nu}$ we found a solution with $e^{-\s}=c^2f^2$, where $f$ is the conformal factor 
and $c$ is a non-zero real constant. The point is that we need to solve $\dd* \th\inv=0$, 
which is typically incompatible with 
$e^{-\s}=const$. 

To work out the difficulties more clearly, consider a given diagonal metric
	$$g=\mbox{diag}(g_{00},g_{11},g_{22},g_{33})$$
and the simple ansatz
	$$\th\inv = \th\inv_{01}\ \dd x^0\ww \dd x^1 + \th\inv_{23}\ \dd x^2\ww \dd x^3,$$
i.e.
	$$\th\inv_{\mu\nu} = 
	\begin{pmatrix}
		0 & {\th_{01}} & 0 & 0 \\ 
  		{-\th_{01}} & 0 & 0 & 0 \\
   		0 & 0 & 0 & {\th_{23}} \\
    		0 & 0 & {-\th_{23}} & 0 \\
	\end{pmatrix}.$$
This can be considered to be a perturbation of (\ref{Minkowski Moyal inv}). We now assume $e^{-\s}=1$ and compute the (inverse) embedding metric:
	\begin{eqnarray*}
		G\inv
		&:=& e^{-\s}\ \th\ g\ \th^T 
		= -(\th\inv)\inv\ g\ (\th\inv)\inv \\
		&=& \begin{pmatrix}
				{g_{11}(\th_{01})^{-2}} & 0 & 0 & 0 \\ 
  				0 & {g_{00}(\th_{01})^{-2}} & 0 & 0 \\
   				0 & 0 & {g_{33}(\th_{23})^{-2}} & 0 \\
    				0 & 0 & 0 & {g_{22}(\th_{23})^{-2}} \\
			\end{pmatrix}.\\
	\end{eqnarray*}
The main observation to be made here is that the tensor $\th$ effectively switches the entries of the metric:
	\begin{eqnarray*}
		g_{00} &\leftrightarrow& g_{11},\\		
		g_{22} &\leftrightarrow& g_{33}.
	\end{eqnarray*}
Requiring also $G\inv=g\inv$ we conclude
	$$\th_{01}=\sqrt{g_{00}g_{11}}, \hspace{.3cm} \th_{23}=\sqrt{g_{22}g_{33}}.$$
However, due to the closedness of $\th\inv$, for this ansatz this is only a valid solution if
	$$\frac{\pa}{\pa{x^0}}(g_{00}g_{11})
	=\frac{\pa}{\pa{x^1}}(g_{00}g_{11})
	=\frac{\pa}{\pa{x^2}}(g_{22}g_{33})
	=\frac{\pa}{\pa{x^3}}(g_{22}g_{33})=0.$$

In this section, we show that this problem can be overcome by introducing (``small'', compactified) extra dimensions,
i.e. by considering spaces with structure $\cM^4 \times \cK$, such that the Poisson tensor relates the 
compact space $\cK$ with the non-compact space-time $\cM^4$. 
This is the idea of split noncommutativity \cite{Steinacker:2011wb}, which is interesting for a variety of reasons. 
Most importantly, there are indeed solutions of the IKKT model with this structure.
Here we initiate a more systematic study of such Poisson structures, and show
in particular that there are in fact solutions for $\theta^{\mu\nu}$ such that $e^\sigma = const$
as desired. We will again consider the case of complexified Poisson structures such that $g = G$,
however analogous considerations should also apply for real Poisson structures.

\subsubsection{$G\sim g$ in higher dimensions}

In the case of higher-dimensional branes, we define  the 
 tensor $\cJ$ generalizing \eq{almost complex structure} as follows
\be
{\cJ^\mu}_\nu = e^{-\frac{n-1}{n}\sigma} \theta^{\mu\mu'} g_{\mu'\nu} 
\ee
so that 
\begin{align}
G^{\mu\nu} = - e^{(\frac 2n -1)\s}\, {(\cJ^2)^{\mu}}_\rho  g^{\rho\nu} \,, \qquad
\det \cJ =  1 .
\end{align}
In particular $G_{\m\n} \sim g_{\m\n}$ is possible only if  $\cJ^2 = -\uno$, 
which in turn implies\footnote{this means that $(\cM,e^{-\frac{n-1}{n}\s} g_{\mu\nu},\omega)$ is an almost-K\"ahler manifold.}
$G^{\mu\nu} = e^{(\frac 2n -1)\s}\,  g^{\mu\nu}$. 
In that case, the following relations hold
\begin{align}
\h &:=\tfrac{1}{4} e^\s \,G^{\m\n}g_{\m\n}
  = \frac{n}{2}\,e^{\frac 2n\s}  \nn\\
    e^{\s} G^{\nu\rho} \theta^{-1}_{\rho\mu} G^{\mu\eta} 
& = e^{-\s} \theta^{\nu\mu'}g_{\mu'\eta'}\theta^{\eta'\eta} g_{\eta\nu'}\theta^{\nu'\eta} 
  =  e^{(1-\frac 2n)\sigma} {(\cJ^2)^\nu}_\mu \theta^{\mu\eta} 
\label{theta-hat}
\end{align}
Therefore if $\cJ^2 \sim \d$, the  equations of motion (\ref{eom}) reduce to 
\begin{align}
\nabla_\m(e^{\s} G^{\m\m'}\th\inv_{\m'\n'} G^{\n\n'}) &=  e^{-\s}\th^{\n\r}\pa_\r \h,  \nn\\
\nabla_\m( e^{(1-\frac 2n)\sigma} \theta^{\mu\nu} ) &= - \frac{n}{2}\,e^{-\s}\th^{\n\r}\pa_\r e^{\frac 2n\s} ,  \nn\\
\nabla_\m \theta^{\mu\nu}   
 &= \(1-\frac 2n - \,e^{-2(1-\frac 2n)\sigma}\) \th^{\n\m}\pa_\m \s .
\label{eom-Gg-reduced}
\end{align}
Using the following identity for Poisson tensors (see (58) in \cite{Steinacker:2010rh})
\be
\nabla_\m(e^{-\s}\th^{\m\n}) = 0, \qquad  \nabla_\m\th^{\m\n} = \th^{\m\n}  \del_\m \s
\ee
this reduces to 
\be
 \(2-\frac 2n - \,e^{-2(1-\frac 2n)\sigma}\) \pa_\m \s =0 .
\ee
This holds identically for $2n=4$ consistent with \eq{D Theta}, and it reduces to $e^{\s} = const$
for $2n\neq 4$.
Therefore symplectic structures with $e^\sigma = const$ and $G \sim g$ are also solutions 
of the equations of motion for $2n\neq 4$. 

To understand better the significance of configurations with $G \sim g$, we recall that 
in the case of the Moyal-Weyl quantum plane, the matrix model action is minimized for 
Poisson tensors which satisfy $G \sim g$, at least for the Euclidean case \cite{Blaschke:2010qj}.

\subsection{Solutions with 4 compactified extra dimensions and $e^{-\sigma} = const$}

Let us now consider an eight-dimensional space, with local coordinates such that the metric takes the following diagonal form:
	$$g=\mbox{diag}\left(g_{00},g_{11},g_{22},g_{33},\frac{1}{g_{00}},\frac{1}{g_{11}},\frac{1}{g_{22}},\frac{1}{g_{33}}\right),
		\hspace{.5 cm} \det g=\prod_{\m=1}^8g_{\m\m}=1.
	$$
For this type of metric, the switch of components
	\begin{eqnarray*}
		g_{00} &\leftrightarrow& g_{44}=\frac{1}{g_{00}},\qquad	\quad
		g_{11} \leftrightarrow g_{55}=\frac{1}{g_{11}},\\		
		g_{22} &\leftrightarrow& g_{66}=\frac{1}{g_{22}},\qquad\quad
		g_{33} \leftrightarrow g_{77}=\frac{1}{g_{33}},
	\end{eqnarray*}
amounts to taking the inverse:
	\begin{eqnarray*}
		g &\leftrightarrow& g\inv.
	\end{eqnarray*}
This is achieved via $\ e^{-\s}\th^{\m\m'}\th^{\n\n'}g_{\m'\n'}$ through the following
2-form
	$$\th\inv_{\m\n}
		= \begin{pmatrix}
				0 & 0 & 0 & 0 & {\ 1\ } & 0 & 0 & 0 \\ 
  				0 & 0 & 0 & 0 & 0 & {\ 1\ } & 0 & 0 \\ 
   				0 & 0 & 0 & 0 & 0 & 0 & {\ 1\ } & 0 \\ 
    				0 & 0 & 0 & 0 & 0 & 0 & 0 & {\ 1\ } \\ 
				-1 & 0 & 0 & 0 & 0 & 0 & 0 & 0 \\ 
				0 & -1 & 0 & 0 & 0 & 0 & 0 & 0 \\ 
				0 & 0 & -1 & 0 & 0 & 0 & 0 & 0 \\ 
				0 & 0 & 0 & -1 & 0 & 0 & 0 & 0 \\ 
			\end{pmatrix}, 
		\hspace{.3 cm} \mbox{det}(\th\inv)=1=\mbox{det}(g),$$
so that $e^{-\s}=1$ as desired.
This 2-form is constant and therefore closed, and in fact exact since
	\be
	\label{Moyal-like solution}	
		\th\inv = \dd x^0\ww\dd \xi^4+\dd x^1\ww\dd \xi^5+\dd x^2\ww\dd \xi^6+\dd x^3\ww\dd \xi^7=\dd \left( \sum_{\n=0}^3x^\n \dd \xi^{\n+4} \right),
	\ee
where $\xi^\mu$ with $\mu=4,5,6,7$ are the coordinate functions of the second block. 
The idea is to use such a structure for the description of physical space-time with compactified extra dimensions,
where  $\mbox{diag}(g_{00},g_{11},g_{22},g_{33})$ correspond to physical space-time and 
$g=\mbox{diag}\left(\frac{1}{g_{00}},\frac{1}{g_{11}},\frac{1}{g_{22}},\frac{1}{g_{33}}\right)$ to the 
small extra dimensions. In this way, it is possible to reconcile the equation of motion for the 
Poisson structure (\ref{eom}) with the requirement of a constant dilaton resp. gauge coupling $e^{-\sigma}$.
The required adaptations for the case of Minkowski signature are obvious and will be illustrated below.

\subsubsection{Radially symmetric Schwarzschild-type metrics}

As a physical application we consider the Schwarzschild metric, which in standard coordinates takes the diagonal form
\be
g_{\mu\nu}=\mbox{diag}\left(-\bar{\k}(r),\frac{1}{\bar{\k}(r)},r^2,r^2\sin^2(\vth)\right), \quad \bar{\k}(r)=1-\frac{r_c}{r},
\label{Schwarzschild-metric}
\ee
where $r_c$ is the Schwarzschild horizon. For our purposes however, spherical coordinates are problematic:
If we would follow the construction described 
above, we would end up with the metric component $g_{77}=\frac{1}{g_{33}}=\frac{1}{r^2\sin^2(\vth)}$, which is singular along 
the  $z$-axis.
A more suitable choice is provided by the isotropic coordinates, where the Schwarzschild metric takes the form 
\begin{align}
ds^2 &= \a(\rho) d(x^0)^2 + \b(\rho)(\sum_{i=1}^3 d(x^i)^2), \nn\\
 & \ \ \a(\rho) =-\frac{\left(1-\frac{r_c}{4\rho}\right)^2}{\left(1+\frac{r_c}{4\rho}\right)^2},
 \quad \b(\rho)=\left(1+\frac{r_c}{4\rho}\right)^4, \quad \rho = \sqrt{\sum_{i=1}^3 d(x^i)^2} \ . \nn
\end{align}
These coordinates are defined for $\rho\geq \frac{r_c}4$, i.e. outside of the horizon.
Notice that the spatial part of metric is now conformally flat. 
By following the procedure introduced in the last section, we now obtain the eight-dimensional metric 
\be
g =\mbox{diag}\left(\a(\rho),\b(\rho),\b(\rho),\b(\rho),-\frac{1}{\a(\rho)},\frac{1}{\b(\rho)},\frac{1}{\b(\rho)},
\frac{1}{\b(\rho)}\right),
\hspace{.3 cm} \mbox{det}(g)=-1,
\label{isotropic-extradim}
\ee 
which is regular outside of the Schwarzschild horizon and in which the Moyal-like 2-form (\ref{Moyal-like solution}) 
is the desired solution for a symplectic structure with $e^{-\s}=1$. 

To see this explicitly,  consider more generally a metric of the form \eq{Schwarzschild-metric}.
The isotropic coordinates are obtained by 
introducing a new radial 
coordinate $\rho(r)$, which is related to $r$ by the differential equation 
$\frac {d \rho}{\rho}=\frac{dr}{r\sqrt{\bar{\k}(r)}}$, and
	\begin{eqnarray*}
		x^0(t) &=& t,\\
		x^1(r,\th,\varphi) &=& \rho(r)\sin(\th)\cos(\varphi),\\
		x^2(r,\th,\varphi) &=& \rho(r)\sin(\th)\sin(\varphi),\\
		x^3(r,\th,\varphi) &=& \rho(r)\cos(\th).
	\end{eqnarray*}
This can be integrated explicitly as
	$$\rho(r)=\rho_c\ \mbox{e}^{\int_{r_c}^r\frac{1}{\varrho_1\sqrt{\k(\varrho_1)}}\dd\varrho_1}.$$
Then consider the following extended metric with 4 extra-dimensions
	$$g=\mbox{diag}\left(-\k(r),\frac{1}{\k(r)},r^2,r^2\sin^2(\vth),\frac{1}{\k(r)},\left(\frac{\rho(r)}{r}\right)^2,\left(\frac{\rho(r)}{r}\right)^2,\left(\frac{\rho(r)}{r}\right)^2\right),$$ 
in coordinates $x^0,r,\vth,\varphi,\xi^4,...,\xi^7$.
It acquires the form \eq{isotropic-extradim} in isotropic coordinates, so that  
the exact 2-form 
	\begin{eqnarray*}
		\th\inv
		&=& \dd \left(i x^0 \dd \xi^4 +  \sum_{\n=1}^3x^\n \dd \xi^{\n+4} \right) \\
		&=& \dd \left( i t\ \dd\xi^4 + \rho(r) \sin(\th)\cos(\varphi)\ \dd\xi^5 + \rho(r) \sin(\th)\sin(\varphi)\ \dd\xi^6 + \rho(r) \cos(\th)\ \dd\xi^7 \right)\\
		&=& i\dd t\ww\dd \xi^4+\ \rho(r) \left( \tfrac{\sin(\th)\cos(\varphi)}{r\sqrt{\k(r)}}\ \dd r\ww \dd \xi^5 + \tfrac{\sin(\th)\sin(\varphi)}{r\sqrt{\k(r)}}\ \dd r\ww \dd \xi^6 + \tfrac{\cos(\th)}{r\sqrt{\k(r)}}\ \dd r\ww \dd \xi^7   \right.\\
		&& +\ \cos(\th)\cos(\varphi)\ \dd\th\ww\dd \xi^5-\sin(\th)\sin(\varphi)\ \dd\varphi\ww\dd \xi^5\\
		&& +\ \cos(\th)\sin(\varphi)\ \dd\th\ww\dd \xi^6+\sin(\th)\cos(\varphi)\ \dd\varphi\ww\dd \xi^6 \\
		&& \left.-\ \vphantom{\tfrac{1}{r\sqrt{\k(r)}}}\sin(\th)\ \dd\th\ww\dd \xi^7 \right)
	\end{eqnarray*}
satisfies 
	$$e^{-\s}=1$$ 
and 
	$$G^{\m\n}:=e^{-\s}\th^{\m\m'}\th^{\n\n'}g_{\m'\n'}=\th^{\m\m'}\th^{\n\n'}g_{\m'\n'}=g^{\m\n}.$$
A notable example of this kind of metric is given by $\k(r)=1-\frac{r_c}{r}+\frac{Q^2}{r^2}$
corresponding to the Reissner–Nordstr\"om metric, which was already discussed in the four-dimensional case. 

Up to now, we now we have not specified the topology of the extra dimensions. 
One possibility is to compactify the additional coordinates at each point, to obtain a torus bundle over 
Schwarzschild space. This will be explained in the example of the de Sitter geometry below.

	\subsubsection{Generalizations}
The above construction can be extended easily to more general metrics.  For example, consider the metric
	$$g=\mbox{diag}\left(g_{00},g_{11},g_{22},g_{33},\frac{-1}{g_{00}},\frac{1}{g_{11}},\frac{1}{g_{22}},\frac{f(x^3, \xi^7)}{g_{33}}\right),
			\hspace{.3 cm} \mbox{det}(g)=-f(x^3,\xi^7).$$
We see that we can freely determine the signature of the second block of the metric
by introducing additional factors of $-1$, or more generally introduce non-vanishing functions $f(x^3,\xi^7)$, 
which might depend on the variables of the position it is to be switched with. Then the form
	$$\th\inv_{\m\n}
		= \begin{pmatrix}
				0 & 0 & 0 & 0 & {\sqrt{-1}} & 0 & 0 & 0 \\ 
  				0 & 0 & 0 & 0 & 0 & {\ 1\ } & 0 & 0 \\ 
   				0 & 0 & 0 & 0 & 0 & 0 & {\ 1\ } & 0 \\ 
    				0 & 0 & 0 & 0 & 0 & 0 & 0 & \sqrt{f(x^3,\xi^7)} \\ 
				-{\sqrt{-1}} & 0 & 0 & 0 & 0 & 0 & 0 & 0 \\ 
				0 & -1 & 0 & 0 & 0 & 0 & 0 & 0 \\ 
				0 & 0 & -1 & 0 & 0 & 0 & 0 & 0 \\ 
				0 & 0 & 0 & -\sqrt{f(x^3,\xi^7)} & 0 & 0 & 0 & 0 \\ 
			\end{pmatrix},$$
	$$\mbox{det}(\th\inv)=-f(x^3,\xi^7)=\mbox{det}(g),$$
is still exact and everything works out as before.

\subsection{Solutions with 2 compactified extra dimensions}

For many metrics of interest, one can easily write down an extension with only 2 compactified extra dimensions
such that $e^{-\sigma} = const$.

\subsubsection{de Sitter space}
As an example consider the  product of de Sitter space \cite{Spradlin:2001pw} with a two-dimensional 
warped torus 
	$$\cM^6=\cM^4_{\rm {dS}}\times T^2.$$ 
In planar coordinates,  the metric of $\cM^4_{\rm {dS}}$ is diagonal with entries
$g_{\mu\nu}=\mbox{diag}(-1,e^{-2t},e^{-2t},e^{-2t})$. We define the metric on the product space as 
	$$g_{\cM^6}=g_{\rm dS}\oplus g_{\rm torus}
=\mbox{diag}\left(-1,e^{-2t},e^{-2t},e^{-2t},\frac{1}{e^{-2t}},\frac{1}{e^{-2t}}\right), 
\hspace{.3 cm} \mbox{det}(g_{\cM^6})=-e^{-2t}.$$
We will denote the additional two coordinate functions of the torus as $\xi_1,\xi_2$. The torus metric
$g_{\rm torus} = e^{2t}\diag(1,1)$ is flat with a 
conformal factor $e^{2t}$,  which depends on the time coordinate of $dS^4$. 
The pleasant aspect of this coordinate system is that the metric only depends on one variable, 
which helps to find a closed form.

An appropriate solution for a Poisson structure with with $e^{-\s}=1$ is
\be
(\th\inv_{\m\n})
		= \begin{pmatrix}
				0 & {i\ e^{-t}} & 0 & 0 & 0 & 0 \\ 
  				{-i\ e^{-t}} & 0 & 0 & 0 & 0 & 0 \\
  				0 & 0 & 0 & 0 & {\ 1\ } & 0 \\
  				0 & 0 & 0 & 0 & 0 & {\ 1\ } \\
  				0 & 0 & -1 & 0 & 0 & 0 \\
  				0 & 0 & 0 & -1 & 0 & 0 \\
			\end{pmatrix}, 
		\hspace{.3 cm} \mbox{det}(\th\inv)=-e^{-2t}=\mbox{det}(g).
\label{desitter-theta-extra}
\ee	
i.e. 
\be
\th\inv  = i\ e^{-t}\ \dd t \ww \dd x + \dd y \ww \dd \xi^1 + \dd z \ww \dd \xi^2 
= -\dd\left( i\ e^{-t}\ \dd x + \xi^1\ \dd y + \xi^2\ \dd z \right).
\label{desitter-form-extra}
\ee
The size of the extra-dimensional torus might seem to depend on the location on $dS^4$, however  
we have not yet defined the range of the periodic torus coordinates $\xi^1,\xi^2$. 
We do this by requiring that all tori should have the same volume,
	\be
	\label{volume condition}	
\int_{T^2}\sqrt{|g_{\rm torus}|}\ \dd\xi^1 \dd\xi^2 \overset{!}{=} 1, \qquad\mbox{so that}\quad
		\int_{T^2}\sqrt{|g_{\cM^6}|}\ \dd\xi^1 \dd\xi^2 = \sqrt{|g_{\rm dS}|}
	\ee
independent of $t$.
This ensures that the action for the lowest Kaluza-Klein modes takes the standard form on 4-dimensional 
de Sitter space.
Using $\sqrt{|g_{\rm torus}|}= e^{2t}$, this condition reads
	$$\int \dd\xi^1 \dd\xi^2 e^{2t}\overset{!}{=} 1.$$
For example, a reasonable choice which satisfies this condition is
	$$\xi^1,\xi^2 \in [0, e^{-t})$$
with the endpoints identified, 
i.e.  $\xi^i \in \mathbb{R}\ \mbox{mod}\ e^{-t}$. This defines a torus $T^2$. We have thus specified a 
metric for the product space $\cM_6=\cM^4_{\rm dS}\times T^2$.
It is clear that this  metric is globally well-defined.
For the symplectic form this is not obvious, because it was defined in planar coordinates which 
do not cover the entire de Sitter space. It would be interesting to study if and how $\theta^{-1}_{\mu\nu}$
can be extended to a globally well-defiend symplectic form.

\section{Conclusion and discussion}

We studied complexified Poisson structures which are compatible with a 
(pseudo-) Riemannian structure, as required in the framework of 
noncommutative emergent gravity within matrix models of Yang-Mills type. We focused on 
a natural class of Poisson tensors for which the physical metric and the 
effective metric of the theory coincide. In 4 dimensions, this means that the corresponding symplectic form must be 
proportional to its Hodge dual. In the case of Minkowski signature, there are no real Poisson structures of this type, 
however complexified symplectic structures which are self-dual up to multiplication with $\pm i$ turn out to satisfy this 
condition. Upon imposing a certain reality constraint, the structure of such complexified symplectic structures
is clarified, and a  normal form is established. Furthermore, an appropriate version of the Darboux theorem 
is established by studying a certain foliation of space-time 
defined by the complexified symplectic structure.
Explicit solutions are given for several examples of physically relevant geometries, including Minkowski, 
Schwarzschild and conformally flat space-times. 

One important aspect is the behavior of a particular
scale parameter $e^\s$ of these structures, which is in general space-dependent and governs the 
effective gauge coupling in emergent NC gravity. The corresponding scalar field is found to vary considerably 
for physically relevant space-times. 
To circumvent this problem, we study analogous structures in 
the higher dimensional case, notably for physical space-times with compactified 
extra dimensions. We show that the problem can be solved in this higher-dimensional setting,
by considering symplectic structures which mix the non-compact physical space-time with the 
compactified extra dimensions. This is illustrated in the case of the Schwarzschild and de Sitter metrics.

The results of this paper help to clarify several issues in emergent NC gravity, and show a general 
strategy how to obtain physically relevant solutions. In particular, the idea of a symplectic structure
which mixes space-time  with small extra dimensions appears to be far-reaching and promising.
It would be very interesting to generalize this to more generic geometries. 
Also, some gaps in the present treatment -- such as 
the extension of  \eq{desitter-form-extra} to the entire de Sitter space -- remain to be filled.
The problems studied in this paper should also be of interest in different contexts.

On the other hand, 
the case of real symplectic structures on geometries with Minkowski signature should also be studied further. 
There are  issues with the causality structure which should be clarified. 
In both cases, extensions with compactified extra dimensions as discussed here 
appear to be very promising towards obtaining realistic geometries
within the framework of emergent NC gravity.

\subsection*{Acknowledgements}

This work was supported by the Austrian Science Fund (FWF) under contract  P21610.



\begin{thebibliography}{99}


\bibitem{Doplicher:1994tu}
  S.~Doplicher, K.~Fredenhagen, J.~E.~Roberts,
  ``The Quantum structure of space-time at the Planck scale and quantum fields,''
  Commun.\ Math.\ Phys.\  {\bf 172 } (1995)  187-220.
  [hep-th/0303037].

\bibitem{Amati:1988tn}
  D.~Amati, M.~Ciafaloni, G.~Veneziano,
  ``Can Space-Time Be Probed Below the String Size?,''
  Phys.\ Lett.\  {\bf B216 } (1989)  41.


\bibitem{Steinacker:2010rh}
Harold Steinacker.
\newblock {``Emergent Geometry and Gravity from Matrix Models: an
  Introduction''}.
\newblock {\em Class. Quant. Grav.}, 27:133001, (2010). arXiv:1003.4134v4
  [hep-th].


\bibitem{Blaschke:2010qj}
Daniel~N. Blaschke and Harold Steinacker.
\newblock {``Curvature and Gravity Actions for Matrix Models II: the case of
  general Poisson structure''}.
\newblock {\em Class. Quant. Grav.}, 27:235019, (2010). arXiv:1007.2729v2
  [hep-th].

\bibitem{madore}
  J.~Madore,
  ``An introduction to noncommutative differential geometry and its physical applications,''
  Lond.\ Math.\ Soc.\ Lect.\ Note Ser.\  {\bf 257 } (2000)  1-371.

\bibitem{Blaschke:2010ye}
Daniel~N. Blaschke and Harold Steinacker.
\newblock {``Schwarzschild Geometry Emerging from Matrix Models''}.
\newblock {\em Class. Quant. Grav.}, 27:185020, (2010). arXiv:1005.0499v1
  [hep-th].


\bibitem{Klammer:2009ku}
Daniela Klammer and Harold Steinacker.
\newblock {``Cosmological solutions of emergent noncommutative gravity''}.
\newblock {\em Phys. Rev. Lett.}, 102:221301, (2009). arXiv:0903.0986v2
  [gr-qc].
 

\bibitem{MR2503498}
R. Fintushel and R. Stern, "Six lectures on four 4-manifolds,"
 in ``Low dimensional topology``,
    {\em IAS/Park City Math. Ser.} Vol {\bf 15},
     265--315 (2009);
 


\bibitem{taubes} 
C. H. Taubes, Math. Res. Lett. 
{\em 13} (2006), no. 4, 557–570 


\bibitem{Steinacker:2011wb}
  H.~Steinacker,
  ``Split noncommutativity and compactified brane solutions in matrix models,''
{\em Prog. Theor. Phys.} Vol. {\bf 126}, No.4;
  [arXiv:1106.6153 [hep-th]].


\bibitem{almost-k-einstein}
K. Sekigawa, L.  Vanhecke, 
``Four-dimensional almost Kähler Einstein manifolds.'' 
{\em Ann. Mat. Pura Appl.} {\bf(4)} 157 (1990), 149–160;
K. Sekigawa, A. Yamada,
``Compact indefinite almost Kähler Einstein manifolds.''
{\em Geom. Dedicata} {\bf  132} (2008), 65–79;
J. Armstrong, 
``An ansatz for almost-Kähler, Einstein 4-manifolds.''
{\em J. Reine Angew. Math.} {\bf 542} (2002), 53–84. 



\bibitem{Arnlind:2011rz}
  J.~Arnlind, G.~Huisken,
  ``On the geometry of K\'ahler-Poisson structures,''
  [arXiv:1103.5862 [math.DG]].

\bibitem{Blaschke:2010rg}
Daniel~N. Blaschke and Harold Steinacker.
\newblock {``Curvature and Gravity Actions for Matrix Models''}.
\newblock {\em Class. Quant. Grav.}, 27:165010, (2010). arXiv:1003.4132v3
  [hep-th].



\bibitem{Ishibashi:1996xs}
  N.~Ishibashi, H.~Kawai, Y.~Kitazawa and A.~Tsuchiya,
  ``A large-N reduced model as superstring,''
  Nucl.\ Phys.\  B {\bf 498} (1997) 467
  [arXiv:hep-th/9612115]


\bibitem{Douglas:2001ba}
  M.R.~Douglas and N.A.~Nekrasov,
  ``Noncommutative Field Theory'',
  Rev.\ Mod.\ Phys.\  {\bf 73} (2001) 977--1029
  {\tt[hep-th/0106048]};
  R.J.~Szabo,
  ``Quantum Field Theory on Noncommutative Spaces'',
  Phys.\ Rept.\  {\bf 378} (2003) 207--299
  {\tt[hep-th/0109162]}.


\bibitem{Steinacker:2008ya}
{  H.~Steinacker,
  ``Covariant Field Equations, Gauge Fields and Conservation Laws from
  Yang-Mills Matrix Models,'',
 {\em JHEP} {\bf 0902}: 044,2009
 {\tt [arXiv:0812.3761 [hep-th]]}. }


\bibitem{Steinacker:2007dq} {  H. Steinacker,  
``Emergent Gravity from Noncommutative Gauge Theory''. 
{\em JHEP} {\bf 12} (2007) 049. {\tt [arXiv:0708.2426v1 (hep-th)]}} 

\bibitem{Sexl:1976pg}
  R.~U.~Sexl and H.~K.~Urbantke,
  ``Relativity Groups, Particles. 
Special Relativity and Relativistic Symmetry in Field and Particle Physics''. Springer, 2000

\bibitem{Ibison:2007dv}
M.~Ibison.
\newblock {``On the Conformal forms of the Robertson-Walker metric''}.
\newblock {\em J. Math. Phys.}, 48:122501, (2007). arXiv:0704.2788v2 [gr-qc];
M.~Ibison.
\newblock {``Static forms of the Robertson-Walker spacetimes''}.
\newblock (2007). arXiv:0704.3265v1 [gr-qc].


\bibitem{Spradlin:2001pw}
Marcus Spradlin, Andrew Strominger, and Anastasia Volovich.
\newblock {``Les Houches lectures on de Sitter space''}.
\newblock (2001). arXiv:hep-th/0110007v2.




\end{thebibliography}
\end{document}